\documentclass[letterpaper]{article} % DO NOT CHANGE THIS
\usepackage{aaai2026}  % DO NOT CHANGE THIS
\usepackage{times}  % DO NOT CHANGE THIS
\usepackage{helvet}  % DO NOT CHANGE THIS
\usepackage{courier}  % DO NOT CHANGE THIS
\usepackage[hyphens]{url}  % DO NOT CHANGE THIS
\usepackage{graphicx} % DO NOT CHANGE THIS
\usepackage{natbib}  % DO NOT CHANGE THIS AND DO NOT ADD ANY OPTIONS TO IT
\usepackage{caption} % DO NOT CHANGE THIS AND DO NOT ADD ANY OPTIONS TO IT
\nocopyright % -- Your paper will not be published if you use this command

\usepackage{algorithm}
\usepackage{algorithmic}

\usepackage{booktabs}
\usepackage{adjustbox}
\usepackage{amsmath}

\usepackage{xcolor}

\definecolor{myyellow}{RGB}{204,153,0}
\definecolor{myblue}{RGB}{66, 133, 244}
\definecolor{mygray}{HTML}{F5F5F5}

\usepackage{array}

\usepackage{newfloat}
\usepackage{tikz}
\usetikzlibrary{shadows.blur}
\usepackage{listings}
\DeclareCaptionStyle{ruled}{labelfont=normalfont,labelsep=colon,strut=off} % DO NOT CHANGE THIS
\floatstyle{ruled}
\newfloat{listing}{tb}{lst}{}
\floatname{listing}{Listing}
\usepackage{times}  % or newtxtext/newtxmath

\usepackage{tabularx}
\usepackage{multirow}
\usepackage{kotex}
\title{Overstating Attitudes, Ignoring Networks: \\LLM Biases in Simulating Misinformation Susceptibility}
\author{
    Eun Cheol Choi\textsuperscript{\rm 1,\rm 2},
    Lindsay E. Young\textsuperscript{\rm 1},
    Emilio Ferrara\textsuperscript{\rm 1,\rm 2}
}
\affiliations{
    \textsuperscript{\rm 1}Annenberg School for Communication and Journalism, University of Southern California\\
    \textsuperscript{\rm 2}Thomas Lord Department of Computer Science, University of Southern California\\
    
    \{euncheol, lindsay.young, emiliofe\}@usc.edu\\
    
}

\begin{document}

\maketitle

\begin{abstract}
Large language models (LLMs) are increasingly used as proxies for human judgment in computational social science, yet their ability to reproduce patterns of misinformation susceptibility remains unclear. We evaluate whether LLM-simulated survey respondents replicate human patterns of misinformation belief and sharing. Using participant profiles from three online surveys that include network, demographic, attitudinal, and behavioral features, we prompt LLMs to simulate survey responses to misinformation items and compare the results to human data on distributions and associations. LLM simulations capture broad distributional tendencies and show modest correlation with human responses. However, they systematically overstate the association between belief and sharing. Linear models fitted to simulated responses show substantially inflated explained variance compared to those fitted to human data. They also place disproportionate weight on attitudinal and behavioral features while largely ignoring personal network characteristics, a pattern not observed in human data. Analyses of LLM training data and model-generated reasoning paths suggest that these distortions reflect systematic biases in how misinformation-related concepts are represented. Our findings indicate that LLM-based survey simulations are better suited for diagnosing systematic deviations from human judgment than for substituting for it.
\end{abstract}

% \begin{links}
%     \link{Code}{https://github.com/EunCheolChoi0123/llm-simulating-misinformation-susceptibility}
% \end{links}

\section{Introduction}

\begin{figure*}[ht]
    \centering
    \includegraphics[width=.98\linewidth]{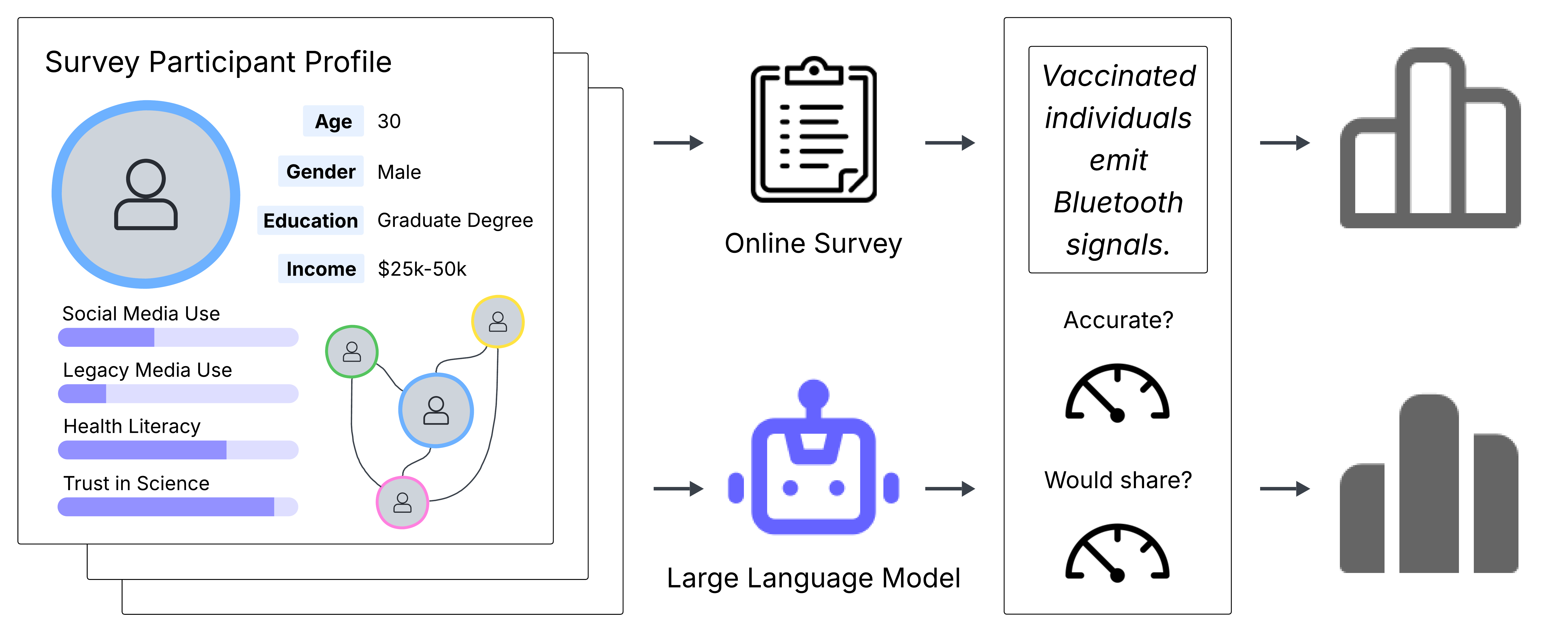}
    \caption{Study pipeline. Human survey participants provided demographic, attitudinal, behavioral, and personal network information through online surveys. Structured profiles containing the same information were provided to large language models (LLMs). Both human participants and LLMs evaluated the same set of false claims across domains and contexts, reporting their beliefs and sharing intentions. We directly compared the resulting outcome distributions and estimated predictor–outcome relationships from both sources, aiming to assess how closely LLM-generated responses mirror human response patterns and to identify systematic differences between simulated and empirical results.}
    \label{fig:pipeline}
\end{figure*}

With the rapid advancement of large language models (LLMs), scholars have increasingly explored their use as simulated agents in social network settings, including scenarios of misinformation spread \cite{pastor2024large, yang2024oasis}. Historically, research focused on simple automated scripts or social bots \cite{ferrara2016rise}. Earlier misinformation research has often turned to agent-based models or epidemic-style diffusion frameworks to simulate how falsehoods spread across populations \cite{tambuscio2019fact, gausen2021can}. While powerful, these approaches have been criticized for limited ecological validity: agents typically follow simplified heuristics, and their behaviors are modeled in stylized rather than psychologically grounded ways \cite{chuang2023simulating}. The appeal of LLMs is that, when conditioned on prompts describing personal traits and contexts, they may generate responses that approximate human judgments, incorporating demographic characteristics, attitudes, and situational cues \cite{tornberg2023simulating, mou2024unveiling, chang2025llms}.

Yet, it remains unclear whether such LLM-based agents recover the empirical relationships between misinformation susceptibility (belief in false claims and willingness to share them) and key social predictors. Prior research with human subjects shows that susceptibility is shaped not only by demographic, cognitive, and attitudinal attributes \cite{nan2022people, ecker2022psychological}, but also by the composition and structure of \textit{personal networks} \cite{facciani2024personal}. If LLM agents fail to reproduce these relationships, simulations of misinformation dynamics may overlook or mischaracterize key drivers. However, only a handful of studies have conducted LLM simulations of misinformation susceptibility \cite{ma2024simulated, pratelli2025evaluating, dash2025persona, plebe2025ll}, and these have focused primarily on sociodemographic predictors while largely neglecting network features.

\subsection{Contributions of This Work}
This study addresses a key gap in LLM-based social simulations by systematically evaluating whether LLMs reproduce patterns associated with both network-based and individual-level factors related to misinformation susceptibility. Using three distinct online survey datasets as ground truth, we directly compare LLM-based survey simulations against human responses to address the following research questions:

\begin{description}
    \item[\textbf{RQ1:}] \textit{To what extent do LLMs approximate the empirical distributions of human misinformation belief and sharing?}
    \item[\textbf{RQ2:}] \textit{To what extent do LLMs recover the empirically observed structure of associations among social predictors and misinformation susceptibility?}
    \item[\textbf{RQ3:}] \textit{Do LLMs systematically differ from human data in the estimated effects of specific features on misinformation susceptibility?}
\end{description}

\section{Related Works}

\subsection{Misinformation Susceptibility} 
A large body of survey research has examined individual differences in misinformation susceptibility. Demographic factors such as age, gender, race/ethnicity, political identification, and education have all been linked to variation in susceptibility \cite{nan2022people, ecker2022psychological}. Cognitive and attitudinal factors, including trust in science, literacy, and trust in media, also influence whether individuals accept or reject misinformation. Network dynamics are also related to susceptibility: discussions with peers can reinforce misinformation through interpersonal validation and repeated exposure \cite{jia2025misinformation}. Thus, it is important to move beyond examining susceptibility in isolation and consider the broader social environment. Network composition has been shown to be a key predictor of both belief in and sharing of false headlines \cite{facciani2024personal}.

Misinformation susceptibility is multi-dimensional: not everyone exposed to misinformation believes or shares it \cite{altay2023misinformation, luceri2025susceptibility}; belief and sharing are tightly correlated, yet they are conceptually and empirically distinct \cite{nan2022people, ecker2022psychological, sirlin2021digital}. Notably, sharing does not always signal belief: one study found that 16\% of headlines were shared despite being judged inaccurate \cite{pennycook2021psychology}. Recent work shows that sharing is often driven less by mistaken beliefs than by social and emotional motives—for example, identity expression, entertainment, or strategic signaling) \cite{paletz2018multidisciplinary, altay2022if, lee2023political}. Accordingly, treating belief and sharing as separate targets and evaluating both is essential to better capture the nuances of misinformation susceptibility.

\subsection{Social Simulation with LLMs} 

Research on social simulation with LLMs spans two broad directions. A first line of work focuses on \textit{multi-agent interactions}, where agents (i) freely interact with one another \cite{park2022social, park2023generative}, (ii) interact within social networks under constraints or restricted action spaces \cite{ashery2025emergent, chang2025llms, chuang2023simulating}, or (iii) specifically operate in social media environments \cite{yang2024oasis, tornberg2023simulating, mou2024unveiling, gao2023s3, orlando2025emergent}. These studies typically explore emergent behaviors \cite{park2023generative, orlando2025emergent} or evaluate how global and temporal distributions of simulated actions compare with empirical traces from social media platforms \cite{yang2024oasis, chang2025llms}. 

A second line of work embeds LLMs in \textit{lab-style experimental or social survey settings}. Here, models are prompted to act as participants in controlled experiments \cite{gui2023challenge}, respond to survey instruments \cite{yu2024large}, or combine both approaches \cite{park2024generative}. Evaluation metrics include alignment with real-life outcome distributions \cite{wang2024not}, correlations among core variables \cite{park2024generative, salecha2024large}, recovery of detailed associations using regression or structural equation models \cite{ma2024simulated, kim2024llm}, and tasks framed as causal inference problems, such as treatment effects and structural recovery \cite{hewitt2024predicting, gui2023challenge, manning2024automated}. Other studies extend this paradigm to forecasting study results \cite{lippert2024can} or automating annotation tasks in NLP research \cite{hu2024quantifying}. 

Despite this promise, several recurring challenges have been identified. One study identified five recurring issues in LLM social simulations: lack of human \textit{diversity}; systematic \textit{biases} in simulated outcomes; \textit{sycophancy}, or user-pleasing outputs; \textit{alienated} outputs that appear accurate but fail to reflect underlying social processes; and limited \textit{generalizability} \cite{anthis2025llm}. Additional concerns include assigning disproportionate weight to predictors that are overrepresented in pretraining corpora \cite{chang2025llms}. Our study contributes by incorporating \textit{diverse} profiles derived from real-world surveys and by systematically examining deviations and potential \textit{biases} in simulated outcomes.

\subsection{LLM-Simulated Misinformation Susceptibility}
A subset of recent research has examined how sociodemographic and psychological factors shape LLM-simulated misinformation susceptibility. For example, one study tested whether “life‐story” profiles encoding demographic features (e.g., age, gender, background traits) influenced simulated judgments of misinformation accuracy and sharing \cite{ma2024simulated}. Another incorporated Big Five personality traits into prompts to evaluate their effects on simulated news discernment \cite{pratelli2025evaluating}. Other studies have incorporated psychological mechanisms and modalities, such as motivated reasoning \cite{dash2025persona} and susceptibility to visual misinformation \cite{plebe2025ll}. One study compared the persuasive impact of misinformation in human--LLM dyads, pairing human participants from diverse demographic backgrounds with LLMs prompted using equally diverse demographic personas \cite{borah2025persuasion}.

These studies suggest that LLM simulations can reproduce certain psychological regularities, but they also raise concerns: models may place disproportionate emphasis on well-documented features while giving less weight to less codified factors. As a result, LLM simulations may be more effective at reproducing established patterns than at revealing novel drivers of susceptibility. Our study extends this line of work by introducing \textit{personal network features}, offering a comprehensive test of whether LLM simulations capture the social bases of misinformation susceptibility.

\section{Methodology}

\subsection{Dataset}

\begin{table}[h]
\centering
\resizebox{0.47\textwidth}{!}{
\begin{tabular}{l c c c c c c}
\toprule
\textbf{Domain} & \textbf{Country} & \textbf{Platform} & \textbf{N} & \textbf{Year} & \textbf{Misinfo} \\
\midrule
Public Health & US & Prolific & 486 & 2023 & 5 items \\
Climate Change & US & Prolific & 377 & 2025 & 3 items \\
Pandemic Politics & KR & Embrain & 708 & 2020 & 5 items \\
\bottomrule
\end{tabular}
}
\caption{Overview of analytic samples from social surveys.}
\label{tab:datasets}
\end{table}

We draw on three survey datasets designed to measure \textit{misinformation susceptibility} alongside a range of \textit{individual- and network-level features} across three domains: public health, climate change, and politically charged issues. Table~\ref{tab:datasets} summarizes the key characteristics of each dataset. 

The \textbf{public health} dataset \cite{choi2026who} was collected online in 2023 from a U.S.-based sample and focuses on five misinformation items related to public health topics, including vaccines and chronic diseases. The \textbf{climate change} dataset \cite{huang2026where, kim2026interpersonal} was collected in 2025 from a U.S.-based online sample as a part of a larger project, and we use the relevant subset of survey modules. This survey focused on misinformation related to climate change and included three image-based stimuli presented as memes. For the purposes of this study, these meme stimuli were transformed into text descriptions prior to analysis. The original images are in Appendix~A. The \textbf{pandemic politics} dataset \cite{ihm2023communication, lee2023role} was collected in South Korea in 2020 using an online panel as a part of a larger project. In this paper, we focus our analysis on five politically charged misinformation items circulating during the COVID-19 pandemic, including claims involving anti-establishment narratives, nationalist or Sinophobic framing, and election-related denial narratives. Across the three studies, estimated hourly compensation was approximately \$10 to \$30, reflecting differences in survey length and payment structures. Average survey completion time ranged from 15 to 20 minutes. These surveys resulted in a total participant compensation cost of approximately \$15{,}000.

Because the three surveys were designed for different research purposes, the sets of measured variables and the scales used to operationalize them differ across datasets. Nevertheless, the variables examined in this study map onto a shared conceptual framework and can be meaningfully grouped into four broad categories. See Appendix~A for the detailed breakdown of the variables.

\begin{itemize}
    \item \textbf{Misinformation susceptibility:} Participants reported their belief in and their willingness to share each of the false claims with others.
    \item \textbf{Personal network measures:} Name-generator and name-interpreter modules captured participants’ discussion partners, alter attributes, and/or alter--alter ties, allowing construction of personal network measures.
    \item \textbf{Demographic measures:} Standard demographic information including age, gender, race/ethnicity, education, income, and/or region.
    \item \textbf{Attitudinal and behavioral measures:} Self-reported attitudes and behaviors, including political leaning, trust in science~\cite{sturgis2021trust}, trust in social media~\cite{obadua2022flow}, traditional media use~\cite{lee2009role}, social media use, and/or health literacy~\cite{chew2008validation}.
\end{itemize}

Together, these measures provide candidates for \textit{multi-dimensional correlates of susceptibility}, incorporating demographic, attitudinal/behavioral, and personal network factors. The dataset serves as an \textit{empirical reference} against which we compare LLM-simulated responses.

\subsection{Experiment Procedure}

\begin{figure}[t]
    \centering
    \begin{tikzpicture}[font=\small]
        \node[
            draw,
            fill=gray!10,
            rounded corners,
            drop shadow={fill=black!30, shadow xshift=3pt, shadow yshift=-3pt},
            inner sep=10pt
        ] {
            \begin{minipage}{0.4\textwidth}
                \raggedright
                \begin{tabular}{l p{0.7\textwidth}}
                    \texttt{SYSTEM} &
                    This survey was conducted in the United States in Oct 2023.\newline
                    Suppose you are a survey participant described below, adopting their demographics, attitudes, behaviors, and personal network.\newline
                    Based ONLY on the profile, answer the given question.\newline
                    If the profile lacks information, make your best estimate from available signals; never reply with “unknown” or similar.\newline
                    Output format (JSON):\newline
                    \{``response'': ``integer''\}\newline
                    
                    \\\
                    \texttt{INPUT} &
                    Participant profile:\newline
                    \{\textit{personal network information}\}\newline
                    \{\textit{demographic information}\}\newline
                    \{\textit{attitudes and behaviors}\}\newline
                    \newline
                    Consider the following claim:\newline
                    ``Vaccinated individuals emit Bluetooth signals.''\newline
                    \newline
                    Question: \newline
                    How would you describe the above claim? (1=Inaccurate, 7=Accurate) \newline
                    \textit{OR} \newline
                    Question: \newline
                    I am willing to share this information with my friends or family. (1=Strongly disagree; 7=Strongly agree) \newline
                \end{tabular}
            \end{minipage}
        };
    \end{tikzpicture}
    \caption{Example prompt format used to simulate public health survey participants’ misinformation belief or sharing intention. Belief and sharing questions were queried separately using identical system prompts, participant profiles, and claims.}
    \label{fig:prompt_survey_sim}
\end{figure}

We prompted large language models (LLMs) to impersonate as \textit{synthetic survey respondents} (Figure~\ref{fig:pipeline}; see Appendix~C for prompts used for other survey domains). Each prompt contained a structured profile describing a survey respondent’s individual attributes and egocentric network features (Figure~\ref{fig:prompt_survey_sim}; see Appendix~D for full profile templates). The profile information was organized into three conceptual blocks: personal network attributes, demographic characteristics, and self-reported attitudes/behaviors. Missing values in survey variables were imputed using median imputation. Models were instructed to answer misinformation belief and sharing intention items based solely on the provided profile, following the same response format as the original human survey.

To account for the stochastic nature of LLM generation and potential sensitivity to prompt design, we implemented two robustness variations: \textit{alternative variable ordering} and \textit{prompting with composite scores}. First, we varied the ordering of profile blocks within the prompt to assess whether response patterns varied with presentation order. Specifically, while the network feature block was presented first in the original prompt format, the alternative ordering placed it last. Second, instead of providing raw item-level responses for each construct, we experimented with supplying composite scores in the profile that summarized each variable, allowing us to assess the robustness of model responses to alternative representations of the same underlying constructs (see Appendix~E for composite score profile templates). Using composite scores in prompting introduces a cost trade-off: while additional preprocessing by researchers shortens prompts and lowers inference costs, it may obscure or omit important contextual information present in the original inputs.

We evaluated a diverse set of LLMs spanning multiple model sizes, architectural families, provider nationalities, levels of content moderation, and degrees of openness. This set included both \textit{reasoning models} and \textit{chat/instruction-tuned models}, reflecting different training objectives and usage paradigms in modern LLM systems. In addition, for chat models, we experimented with explicit chain-of-thought elicitation by adding a dedicated reasoning field to the JSON output to examine whether encouraging intermediate reasoning altered response distributions or the reliance on variables. Models were accessed through official provider APIs or standardized inference services, depending on availability. 

\subsection{Analytic Procedure}

For each respondent profile, we generated simulated survey responses using LLMs and directly compared them to the human survey responses conditioned on the same profile information, enabling profile-level comparisons between simulated and empirical outcomes. We assessed whether LLM outputs (i) reproduced the distribution of misinformation belief and sharing, (ii) approximated the observed relationships between profile features and misinformation susceptibility measures, and (iii) exhibited systematic bias patterns across subgroups.

Our analysis proceeds in six steps to evaluate how closely LLM-simulated survey responses align with human data and identify potential sources of systematic divergence.

\begin{enumerate}
    \item \textbf{Distributional Similarities (RQ1).} 
    We compare the distributions of misinformation belief and sharing between ground truth (human responses) and LLM-simulated results using Jensen--Shannon divergence (JSD) and earth mover’s distance (EMD). 

    \item \textbf{Correlation between Ground Truth and Simulated Responses (RQ1).} 
    We compute ranked correlations (Spearman’s $\rho$) between ground truth and simulated responses to quantify the extent to which LLM-generated outputs align with human responses.

    \item \textbf{Correlation between Simulated Belief and Sharing. (RQ2)} 
    We assess whether LLM simulations exhibit stronger or weaker associations between misinformation belief and sharing than in human data. Although these constructs are strongly related, they remain theoretically and empirically distinct. 

    \item \textbf{Predictive Modeling of Simulated Susceptibility. (RQ2)} 
    We fit Elastic Net regression models to predict simulated misinformation belief and sharing outputs from demographic, cognitive, behavioral, and network features. Elastic Net regularization allows us to account for multicollinearity across correlated predictors while retaining interpretability at the block and variable levels. This analysis enables us to (i) assess the relative predictive contribution of predictor blocks, (ii) evaluate whether simulated outputs place disproportionate predictive weight on certain types of features, and (iii) compare the magnitude and direction of effects against ground-truth models. 

    \item \textbf{Simulation Bias in Predictor Effects (RQ3).} 
    To test whether LLM outputs systematically differ in the estimated relationships among profile features and misinformation outputs, we estimate pooled Elastic Net models that include an indicator for simulated responses and their interactions with key predictors. Interaction terms capture whether the association between a given predictor and misinformation outcomes differs in magnitude or direction between LLM-simulated and human responses.

    \item \textbf{Model Reasoning and Training Corpus Analysis (RQ3).}
    Finally, we conduct exploratory analyses of LLM chain-of-thought processes and training data to explore possible contributors to systematic patterns in simulated responses. For chat-oriented models, we analyze chain-of-thought (CoT) rationales generated alongside survey responses to examine how demographic, attitudinal, behavioral, and network variables are invoked and linked to misinformation belief and sharing decisions. In parallel, we analyze an open-source training corpus using OLMoTrace \cite{liu2025olmotrace}, a tool developed by the Allen Institute for AI (AI2) that traces relevant spans in the pre- and post-training corpus. This allows us to examine how variables of interest co-occur and are framed in the training data. Together, these analyses provide preliminary descriptive evidence regarding whether observed simulation patterns co-occur with model reasoning traces or patterns present in pre- and post-training data.

\end{enumerate}

\section{Results}

Descriptive statistics for variables derived from human participants’ responses are reported in Appendix~A. The rank correlations (Spearman’s $\rho$) between misinformation belief and sharing were $.606$ for public health, $.418$ for climate change, and $.342$ for political misinformation (Table~\ref{tab:correlation}), consistent with the prior works that while the two susceptibility measures are closely related, they remain empirically distinct \cite{pennycook2021psychology}. Elastic Net models fitted to profile features as predictors and misinformation belief and sharing as outcomes explained modest variance (cross-validated R$^2$ ranging from .042 to .228; Table~\ref{tab:r2}), suggesting that demographic, attitudinal, behavioral, and network characteristics are associated with misinformation belief and sharing, albeit with substantial noise, as is typical in social science data.

\begin{description}
    \item[\textbf{RQ1:}] \textit{To what extent do LLMs approximate the empirical distributions of human misinformation belief and sharing?}
    \item \textsl{LLM outputs exhibit partial alignment with broad distributional patterns of human misinformation susceptibility and exhibit moderate correlation with human responses. However, distributional similarity and correlation strength vary across models, prompting formats, and survey contexts, with no single configuration consistently performing best.}
\end{description}

To answer RQ1, we computed distributional divergence and rank-order agreement between ground-truth and LLM-simulated misinformation susceptibility across survey domains, model architectures, and prompting formats (Appendix~B). Across most of the settings, LLM-simulated outputs are moderately correlated with human responses, though the degree of divergence and correlation varies by domain, model, and prompt configuration, and no single model or prompting strategy consistently outperforms others across metrics.

Given the absence of a dominant configuration, and to reduce model-specific noise while preserving consistent differences observed across prompting formats, we aggregate LLM-simulated outputs at the prompting-format level in subsequent analyses, rather than identifying a best-performing system. For each survey domain, we average misinformation measures across all models within each prompting format, yielding three aggregated outputs per belief and sharing outcome.

\begin{description}
    \item[\textbf{RQ2:}] \textit{To what extent do LLMs recover the empirically observed structure of associations among social predictors and misinformation susceptibility?}
\end{description}

\begin{description}
    \item \textsl{Although LLM outputs capture broad associations between profile features and misinformation outcomes, they exhibit consistent deviations from empirical patterns, including (i) an exaggerated coupling between belief and sharing, (ii) inflated explained variance, and (iii) disproportionate emphasis on attitudinal and behavioral predictors while underweighting network-based factors.}
\end{description}

\begin{table}[b]
\centering
\resizebox{.47\textwidth}{!}{
\begin{tabular}{lccc}
\toprule
 & Health & Climate & Political \\
\midrule
Ground Truth   & .606 & .418 & .342 \\
\midrule
LLM-Simulated \\
\addlinespace
\quad Original Prompt  & .925 & .956 & .749\\
\quad Alternative Ordering  & .930 & .957 & .773 \\
\quad Composite-Score & .898 & .953 & .742 \\
\bottomrule
\end{tabular}
}
\caption{Rank correlation ($\rho$) between misinformation belief and sharing across domains.}
\label{tab:correlation}
\end{table}

\begin{figure*}[t]
    \centering
    \includegraphics[width=.75\textwidth]{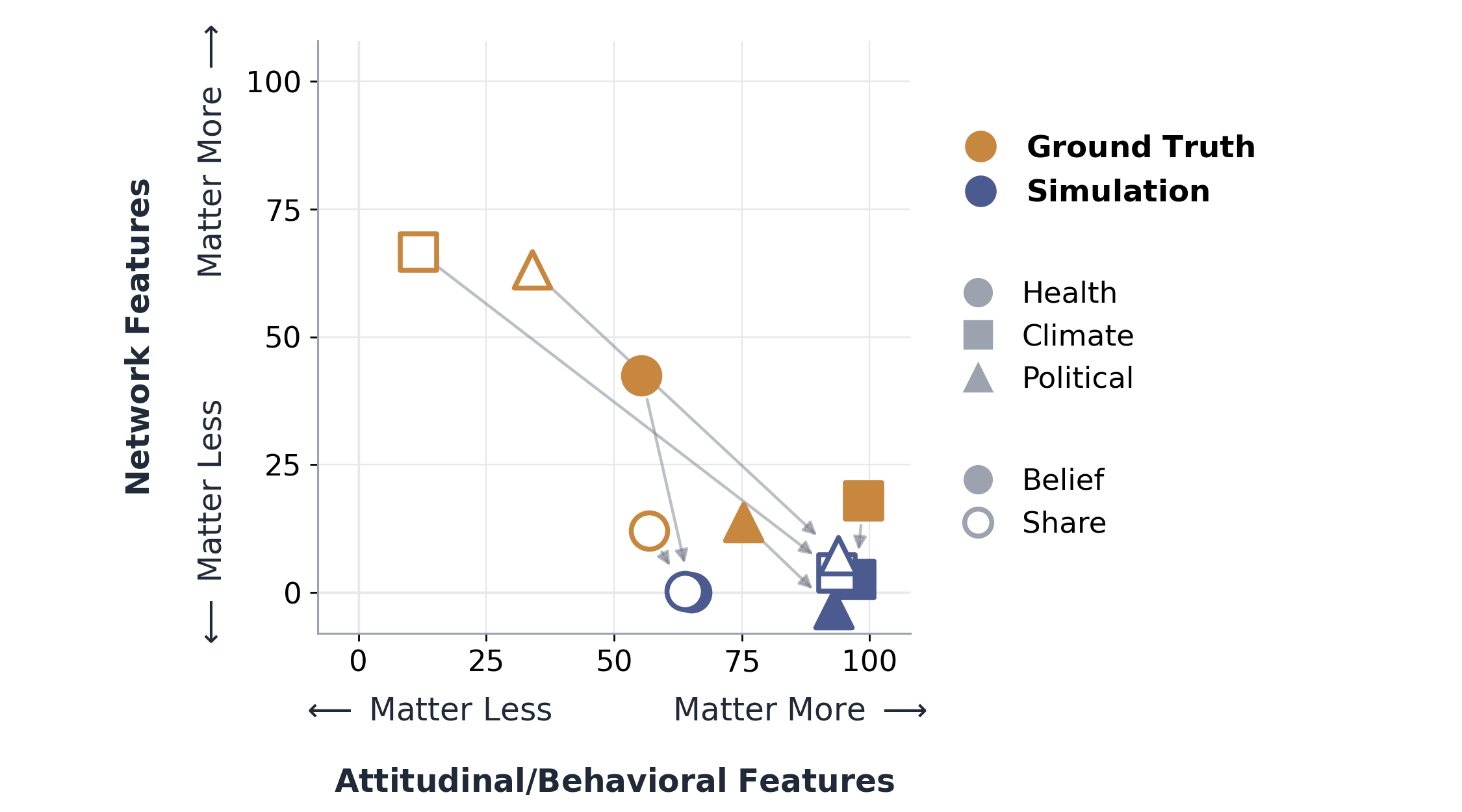}
    \caption{Simulation distorts the structure of feature importance relative to ground truth. Each point shows the proportion of predictive power lost when a feature block is removed: $1 - R^2_{\mathrm{block\ removed}}/R^2_{\mathrm{full\ model}}$ (in \%); higher values indicate that a block matters more. The x-axis shows the importance of attitudinal/behavioral feature blocks and the y-axis the importance of network feature blocks, computed separately for \textcolor{myyellow}{ground-truth} and \textcolor{myblue}{simulated} outcomes. Arrows connect each ground-truth point to its simulated counterpart. Values are averaged across three prompt specifications, and all $R^2$ values are five-fold cross-validated.}
    \label{fig:block_removal}
\end{figure*}

First, we focus on the correlation between two forms of misinformation susceptibility, belief and sharing. As shown in Table~\ref{tab:correlation}, LLM-simulated responses exhibit near-monotonic correlations between belief and sharing across domains, substantially exceeding those observed in human data. Thus, LLM outputs treat belief and sharing as much more tightly coupled than human responses do, attenuating the empirical dissociation between the two measures.

Next, we examine the extent to which LLM outputs reflect the empirical structure of associations linking profile features to misinformation susceptibility. To this end, we fit Elastic Net regression models predicting both misinformation belief and sharing from demographic, attitudinal/behavioral, and network variables, and compare models trained on LLM-simulated outputs with those trained on human ground-truth responses. Across all domains, models fitted on LLM-simulated responses achieve substantially higher $R^2$ values than those fitted on human data, often by an order of magnitude (Table~\ref{tab:r2}). This pattern indicates that simulated responses exhibit much stronger alignment with the provided profile features than is observed in human data, yielding deterministic mappings uncharacteristic of human belief or behavior.

\begin{table}[h]
\centering
\resizebox{.47\textwidth}{!}{
\begin{tabular}{lcccccc}
\toprule
 & \multicolumn{2}{c}{Health} & \multicolumn{2}{c}{Climate} & \multicolumn{2}{c}{Political} \\
\cmidrule(lr){2-3} \cmidrule(lr){4-5} \cmidrule(lr){6-7}
 & Belief & Sharing & Belief & Sharing & Belief & Sharing \\
\midrule
\addlinespace
Ground-Truth & .069 & .191 & .102 & .205 & .228 & .042 \\
% \addlinespace
\midrule
\addlinespace
LLM-Simulated \\
\addlinespace
\quad Original & .831 & .862 & .759 & .752 & .740 & .580 \\
\quad Alt. Order & .810 & .874 & .741 & .735 & .751 & .592 \\
\quad Composite & .687 & .808 & .722 & .734 & .736 & .703 \\
\bottomrule
\end{tabular}
}
\caption{Five-fold cross-validated out-of-sample explained variance ($R^2$) for ElasticNet regression models across domains and outcomes. Models predict misinformation beliefs or sharing intentions from human survey data or LLM-simulated responses, using demographic, attitudinal/behavioral, and network features.}
\label{tab:r2}
\end{table}

Figure~\ref{fig:block_removal} examines the contribution of each feature block by removing blocks from the full model and measuring the resulting drop in explained variance. This procedure serves as an ablation-like check, isolating the extent to which each feature block contributes to the predictive signal. In human data, both attitudinal/behavioral and network features carry substantial predictive signal for belief and sharing outcomes across all domains, indicating that misinformation susceptibility reflects both individual dispositions and network context.

LLM-simulated data, however, provide a systematically distorted picture. Attitudinal/behavioral features dominate prediction in every condition, while removing network features produces little or no loss relative to the full model; a pattern absent in human data. Simulated responses thus exhibit an individual-centric structure: the simulation \textit{overstates} the role of attitudes and behaviors in shaping susceptibility while largely \textit{ignoring} the social network. Full per-prompt results are reported in Appendix~F.

\begin{description}
    \item[\textbf{RQ3:}] \textit{Do LLMs systematically differ from human data in the estimated effects of specific features on misinformation susceptibility?}
    \item \textsl{LLM-simulated responses exhibit structured differences in feature effect estimates, with consistently larger deviations for salient attitudinal and behavioral predictors such as trust in science, political leaning, and social media use, relative to human data. Exploratory analyses of model reasoning traces and open-source pre-/post-training corpus indicate that these predictors are frequently emphasized in both reasoning outputs and training data contexts, which may help contextualize the observed differences.}
\end{description}

To address RQ3, we directly test whether LLM responses systematically over- or under-estimate susceptibility associated with specific profile features. We concatenate human and LLM-simulated responses into a pooled dataset and include an indicator variable denoting simulated observations. For each domain and outcome, we fit Elastic Net regression models that include main effects for all predictors, the simulation indicator, and interaction terms between the simulation indicator and each predictor. These interaction terms capture whether the estimated association between a given feature and misinformation belief or sharing differs between simulated and human data, holding all other features constant.

\begin{figure*}[t]
    \centering
    \includegraphics[width=.95\textwidth]{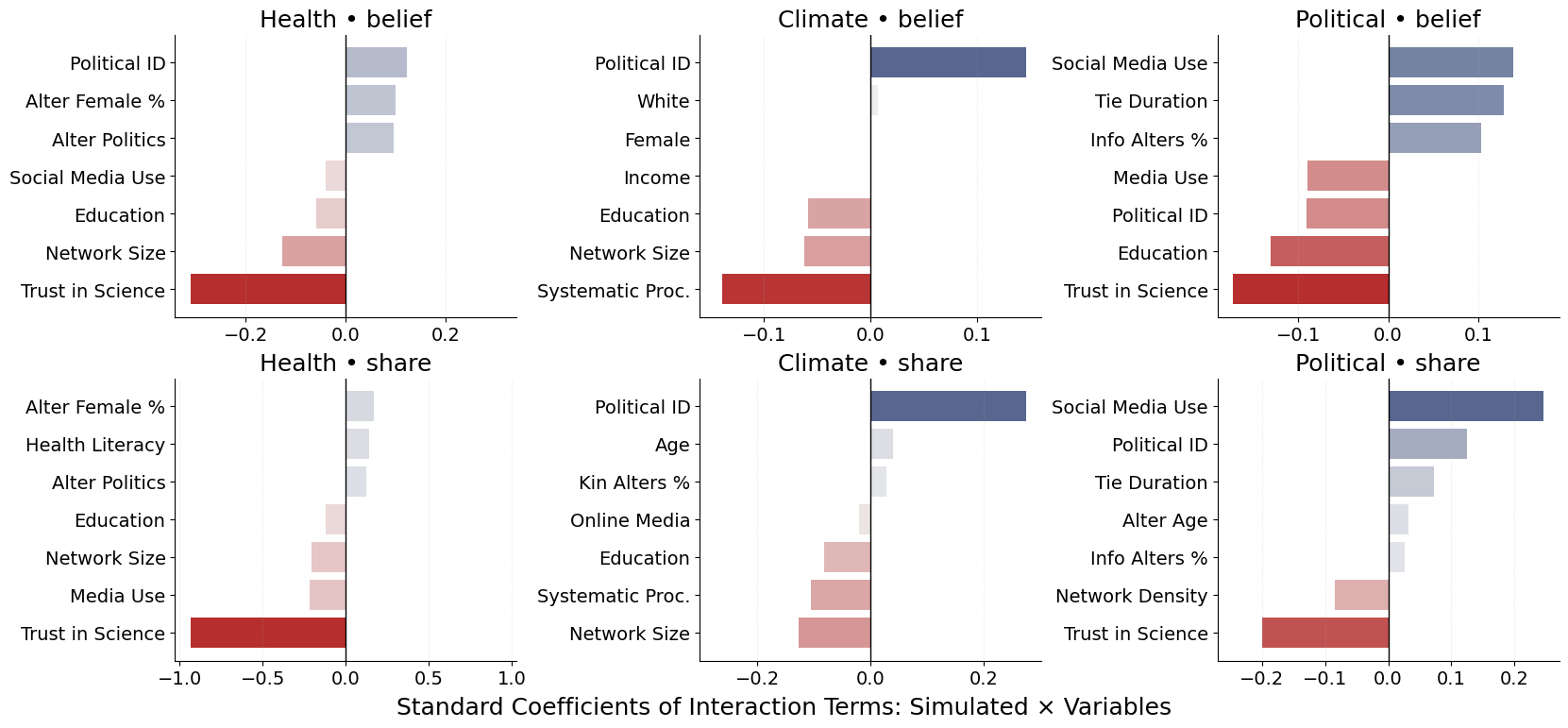}
    \caption{Standardized coefficients for simulation-by-feature interaction terms (Simulated $\times$ Variables) across domains and outcomes. Shown are the seven variables with the largest absolute interaction coefficients. Each coefficient captures the difference in the estimated feature–outcome association between the simulated and ground-truth data (i.e., the change in slope from the simulated to the ground-truth data). The vertical line at zero indicates that the simulated and ground-truth estimates are identical.}
    \label{fig:sim_interactions}
\end{figure*}

Figure~\ref{fig:sim_interactions} reports standardized coefficients for the simulation-by-feature interaction terms. Across domains and outcomes, the largest interaction effects in absolute magnitude are concentrated among attitudinal and behavioral features, including trust in science, political leaning, and social media use, which are commonly cited predictors of misinformation susceptibility in prior literature \cite{nan2022people}. Overall, these patterns suggest that deviations between simulated and human responses are not evenly distributed across predictors, but are more pronounced for a subset of individual-level attitudinal and behavioral features. Relative to human data, simulated responses tend to place greater weight on these salient predictors, while differences associated with network features are generally smaller in magnitude.

To contextualize the patterns observed in RQ3, we conducted post hoc exploratory analyses of the model's reasoning and training data. We first examine model reasoning processes by analyzing chain-of-thought (CoT) rationales generated during misinformation judgments. Our dataset includes 42,583 unique reasoning chains generated by both open-weight reasoning models and chat models prompted with explicit CoT instructions. We then examine whether related regularities are observable in LLM training data. While the training corpora of major proprietary models are not publicly accessible, we analyze open-source pre- and post-training data using OLMoTrace \cite{liu2025olmotrace}, a tool that retrieves up to ten training-data spans most relevant to a given query. Using this approach, we retrieve 187 unique text spans that explicitly reference relationships between misinformation and profile features. The full analytical procedures are documented in Appendix~G.

\begin{figure}[t]
    \centering
    \includegraphics[width=.48\textwidth]{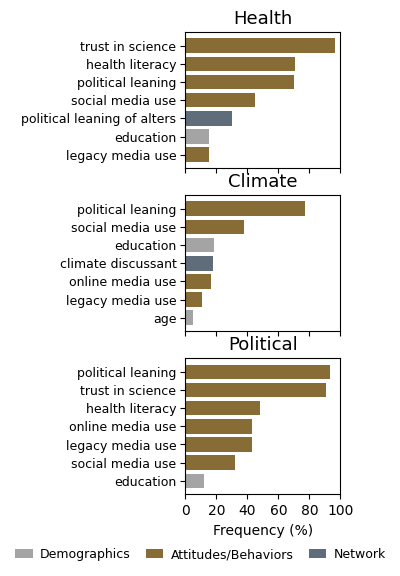}
    \caption{Frequency of the seven most frequently referred variables across reasoning chains by domain.}
    \label{fig:reasoning_analysis}
\end{figure}

\begin{figure}[t]
    \centering
    \includegraphics[width=.48 \textwidth]{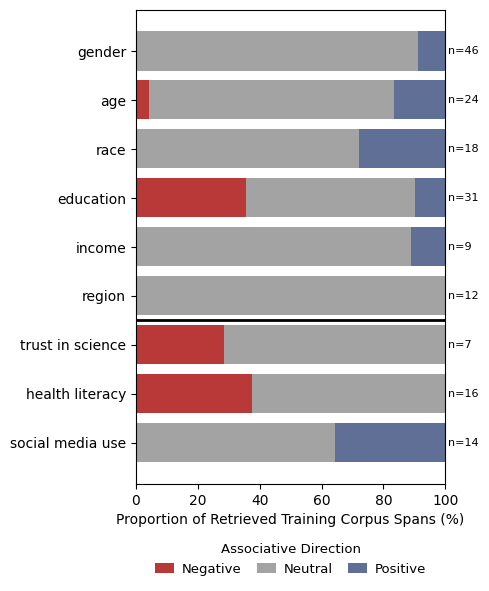}
    \caption{Distribution of inferred association directions in retrieved OLMo pre- and post-training corpus spans. Stacked bars show the proportion of spans classified as indicating negative, neutral, or positive relationships between each variable and misinformation belief or sharing; \textit{n} indicates the number of retained spans per variable. Variables with fewer than five retained spans are omitted.}
    \label{fig:olmotrace_analysis}
\end{figure}

Descriptive analyses of reasoning traces and open-source training corpus point to similar patterns. LLMs frequently invoke attitudinal and behavioral predictors in their reasoning traces, most notably trust in science, political leaning, social media use, and health literacy, while network-related attributes appear less frequently (Figure~\ref{fig:reasoning_analysis}). Furthermore, analysis of the training corpus shows that features such as trust in science, social media use, and health literacy frequently co-occur with misinformation-related concepts in non-uniform, directional patterns (Figure~\ref{fig:olmotrace_analysis}).

\begin{figure*}[t]
    \centering
    \includegraphics[width=.8\linewidth]{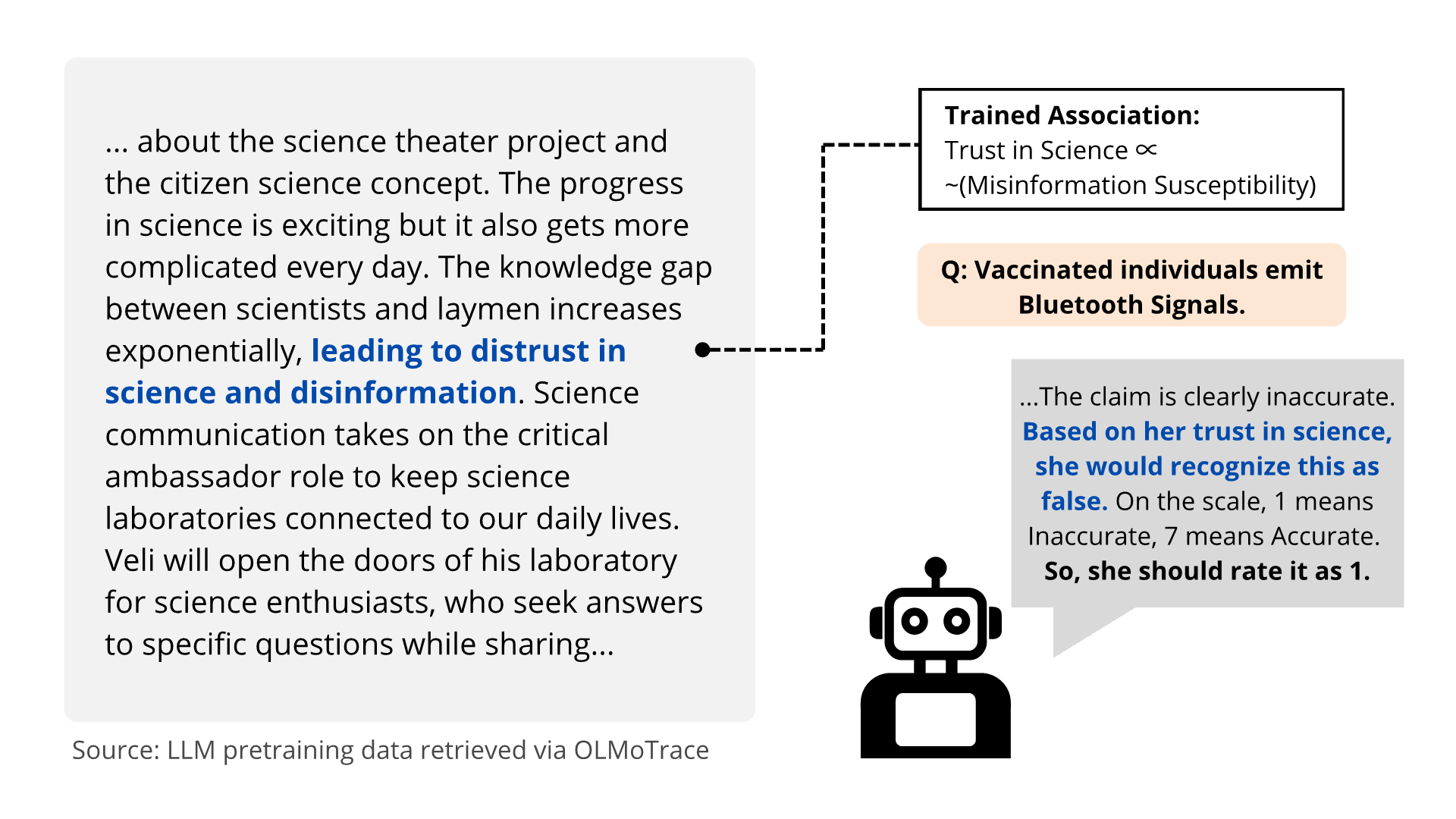}
    \caption{Conceptual illustration of a possible pathway linking training data associations, model reasoning traces, and simulated outputs. The figure illustrates how associations observed in pre- and post-training corpus spans may align with patterns in model-generated reasoning traces, helping to contextualize the exaggerated feature effects and structured distortions observed in LLM-simulated misinformation susceptibility across domains and prompting formats.}
    \label{fig:hypothesized_mechanism}
\end{figure*}

\section{Discussion}

This study evaluates how well large language models (LLMs) approximate patterns of human susceptibility to misinformation when prompted with rich survey profiles that include demographic, attitudinal, behavioral, and personal network information. Across three domains and multiple prompting formats, we find that LLMs reproduce coarse-grained distributional patterns in misinformation belief and sharing and generate responses that are moderately correlated with human judgments. At the same time, substantial and systematic discrepancies emerge beyond this surface alignment. Taken together, these results indicate that LLM-based simulations may be informative for exploratory or diagnostic purposes \cite{dillion2023can, wu2025llm}, but they are limited as stand-ins for human data, especially when the relational structure of social behavior is of central interest.

Our findings reveal systematic divergences that limit the validity of LLMs as substitutes for human data. First, LLM responses substantially exaggerate the association between misinformation belief and sharing. As a result, these two theoretically and empirically distinct dimensions of susceptibility appear far more tightly coupled in simulated data than in human data. In human data, belief and sharing are correlated but separable, reflecting the fact that individuals may share content for social, emotional, or strategic reasons even when they do not fully believe it \cite{ecker2022psychological}. By reducing this differentiation, LLM simulations may obscure distinctions that are empirically present in human behavior, with implications for how misinformation dynamics are modeled in downstream analyses.

Second, regression models fitted on LLM-simulated outcomes exhibit dramatically inflated explained variance compared to models fitted on human responses. This pattern indicates that LLM-generated responses are more tightly aligned with the provided profile features than is observed in human data. Rather than reflecting the substantial variability and residual noise that characterize human beliefs and behaviors, simulated responses appear comparatively more regular and predictable given the same inputs \cite{kaiser2025simulating, anthis2025llm}. As a result, the relationships estimated from simulated data may overstate the degree to which observed features account for misinformation belief and sharing in real populations.

Crucially, this over-structuring is not uniform across predictors. In human data, both attitudinal/behavioral and personal network features contribute meaningfully to explaining misinformation susceptibility. In contrast, LLM simulations place substantially greater weight on attitudinal and behavioral variables, while assigning comparatively less explanatory importance to network features. As a result, LLM outputs tend to yield larger estimated associations for well-documented attitudinal and behavioral predictors, accentuating familiar individual-level explanations, while producing weaker or less stable associations for less codified network-related factors.

Our exploratory analyses of model reasoning and training data provide suggestive evidence regarding patterns that co-occur with the observed distortions. LLMs more frequently reference attitudinal and behavioral variables when generating reasoning traces for susceptibility judgments, with political leaning and trust in science referenced most frequently across domains. By comparison, network-related attributes are referenced relatively infrequently in these reasoning traces. Taken descriptively, this pattern indicates that LLM-generated responses are more often accompanied by references to salient individual-level predictors that are prominent in prior misinformation research \cite{nan2022people}.

Complementing this observation, our analysis of an open-source training corpus shows that several of these same attitudinal and behavioral variables frequently co-occur with misinformation-related concepts in the retrieved text spans. While these analyses do not establish a causal link, the convergence between reasoning traces and training data resonates with concerns about LLMs’ reliance on statistical stereotypes overrepresented in their training data \cite{kotek2023gender, anthis2025llm}. More broadly, these findings align with ongoing discussions about fairness, representation, and the potential for generative AI systems to reproduce societal stereotypes \cite{ferrara2024fairness}.

Taken together, our study suggests that LLM-based survey simulations do not simply differ from human responses due to random variation. Instead, they exhibit patterned deviations that are consistent with differences observed in both reasoning traces and training data associations. In particular, LLM outputs tend to be more strongly associated with features that are well documented and frequently discussed in existing literature, while placing comparatively less emphasis on relational and contextual factors that are less explicitly represented in a text-based corpus. This imbalance is consistent with prior observations that deep neural network models may rely more heavily on readily available or salient cues \cite{geirhos2020shortcut}. 

Unlike attitudes or demographics, personal network features are inherently relational, higher-order, and often non-narrative. They require combining information across multiple entities and ties rather than mapping a single attribute to an outcome, a representational challenge that has been shown to benefit from explicit relational inductive biases \cite{battaglia2018relational}. This representational mismatch may help explain why network features appear less influential in simulated outputs, even when such information is available in the prompt. Consistent with this interpretation, our robustness checks indicate that surface-level prompt modifications alone are unlikely to substantially alter these patterns. Addressing such limitations may therefore require architectural or representational approaches that more directly encode relational structure, rather than relying exclusively on text-based descriptions \cite{tang2024graphgpt, chen2024llaga}.

Our results have important implications for the use of LLMs in social simulation. While LLMs may be useful for examining how well-established predictors are reflected in model outputs, they should not be treated as interchangeable proxies for human respondents, particularly in studies where social context and network structure play a central role. In such settings, reliance on LLM-based simulations may yield overly regularized relationships that place disproportionate emphasis on individual-level factors while underrepresenting social interaction, diffusion, and peer influence. More broadly, our work highlights the need for evaluation frameworks that move beyond surface-level distributional alignment and explicitly test whether simulated relationships preserve key structural properties of human social processes. Without such evaluation, LLM simulations risk implicitly encoding simplified or normative assumptions about how social behavior works. 

Methodologically, this study provides a generalizable pipeline for diagnosing bias and distortion in LLM-based social surveys, combining distributional metrics, predictive modeling, interaction analysis, reasoning inspection, and training-data tracing. Future work should extend this approach to additional domains, alternative network representations, and emerging model architectures, and examine whether targeted training, prompting, or architectural interventions can reduce the systematic patterns documented here.

\subsection{Limitations}
This study has several limitations that warrant caution in interpreting the results. Although we draw on three survey datasets spanning different domains and national contexts, they rely on online panels, which may limit representativeness and introduce sampling biases. While we aimed to cover multiple domains and items, the scope of misinformation topics remains necessarily constrained, and results may not extend to other issue areas, content formats, or claim types. The climate survey included image-based misinformation stimuli that were converted to text for this study, and the political misinformation survey was administered in Korean and translated for analysis, raising the possibility that modality- or language-specific cues relevant to human judgment were attenuated or altered. Future work should extend LLM-based simulations to multimodal settings \cite{plebe2025ll} and multilingual contexts to more closely approximate how misinformation is encountered and evaluated in real-world settings. 

In addition, our interpretation of model distortions, such as the inflated influence of trust in science or political leaning, is primarily descriptive. While analyses of reasoning traces and training corpus provide convergent but exploratory evidence for potential sources of these patterns, they do not establish a causal relationship between training data, reasoning processes, and simulated response behavior. Stronger theoretical and empirical frameworks will be necessary to explain why such biases arise and under what conditions they may be mitigated. Taken together, these limitations underscore that our findings should be viewed as an initial diagnostic rather than a comprehensive or definitive account, and that LLM-based survey simulations should be interpreted cautiously and used in conjunction with diverse, representative, and theoretically grounded human data.

\section{Ethical Statement}
All human data were collected through an online survey of U.S. adults, with approval and oversight from the university’s institutional review board, in accordance with standard ethical guidelines for human-subjects research. Participation was voluntary, responses were anonymized, and no personally identifiable information was retained; all analyses were conducted at the aggregate level to further minimize risk. Thus, we consider the likelihood of harmful consequences from using these data to be low. 

We acknowledge the possibility of mischaracterizing certain groups’ susceptibility to misinformation and, if such patterns were taken at face value, of reinforcing bias in downstream simulation studies. A central motivation of this work is to make such risks visible by demonstrating that surface-level alignment between real and simulated responses can mask important discrepancies in relational structure and feature effects. This is a challenge that future studies in this area will also need to confront.

\section*{Acknowledgments} 
This project was in part supported by the NSF (Award Number 2331722).

\textbf{AI Tools Usage Disclosure}. During the preparation of this work, the authors used ChatGPT for text refinement and code review. The authors reviewed and edited the content as needed and take full responsibility for the publication's content.

\bibliography{aaai2026}

\clearpage

\appendix\section{Appendix A. Survey Materials}
\appendix\subsection{A.1. Misinformation Items}

\begin{table}[h]
\centering
\begin{tabular}{@{}p{0.14\linewidth}p{0.85\linewidth}@{}}
\toprule
\textbf{Survey}   & \textbf{Claim} \\ \midrule
 & 1. Ivermectin pills, known as the antiparasitic drug, have been approved by the FDA to treat COVID-19\\ 
  & 2. Vaccinated individuals emit Bluetooth signals\\ 
Public Health   & 3. Most generic sunscreens on the market contain benzenes which are a cancer-causing agent\\
 & 4. Diabetes can be treated by wearing a copper bracelet\\
  & 5. WHO has said smoking prevents people from getting infected with the novel coronavirus\\
\midrule
 & 1. The increase in the global polar bear population from about 5,000 in the 1960s to over 25,000 today proves that global warming is exaggerated or a hoax. \\
Climate Change  & 2. The Paris Climate Treaty hurts the U.S. while letting China and India pollute more, making it useless for protecting the environment. \\
   & 3. A single eruption of Mount Etna releases more carbon dioxide than all human activity combined. \\
\midrule
 & 1. The government is deliberately minimizing the group of people eligible for COVID-19 diagnostic test to reduce the number of confirmed cases before the election. \\
 & 2. The government has exclusive control over COVID-19 clinical information, refusing to share them with the experts. \\
Pandemic Politics   & 3. Purchasing public masks at a pharmacy will lead to leakage of personal information which will be used in election fraud. \\
 & 4. There was a shortage in mask at hospitals in Korea because the government sent masks to China. \\
  & 5. The government is providing masks purchased with tax money to China. \\ 
\bottomrule
\end{tabular}
\caption{Misinformation items used in the study.}
\label{tab:claims}
\end{table}

\begin{figure}[H]
    \centering
    \includegraphics[width=.4\textwidth]{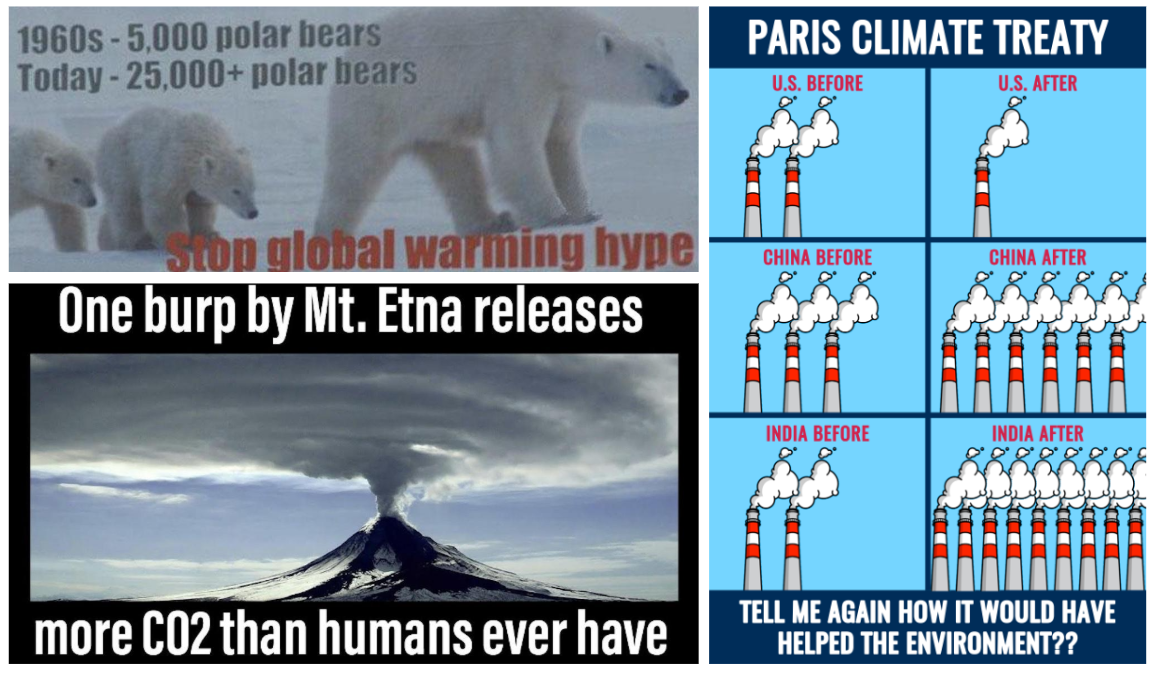}
    \caption{Three meme stimuli used in climate change survey, namely (i) stop global warming hype (top left), (ii) Paris Climate Treaty (right), and (iii) Mount Ena CO2 (bottom left).}
    \label{fig:climate_meme}
\end{figure}

\pagebreak
\appendix\subsection{A.2. Descriptive Statistics}

\begin{table}[h]
\centering
\resizebox{0.4\textwidth}{!}{
\begin{tabular}{ll}
\toprule
\textbf{Variables} & \textbf{$M$ (SD) or \%} \\ 
\midrule
\multicolumn{2}{l}{\textbf{Demographics}} \\
Female & 50.82\% \\
White & 76.95\% \\
Age & 45.45 (16.01) \\
Education & 4.24 (1.31) \\
Income & 3.23 (1.53) \\
\midrule
\multicolumn{2}{l}{\textbf{Attitudinal/Behavioral}} \\
Political identification & 3.29 (1.78) \\
Trust in science & 3.22 (0.65) \\
Social media use & 2.79 (1.74) \\
Health literacy & 4.39 (0.68) \\
Health media exposure & 1.75 (0.78) \\
\midrule
\multicolumn{2}{l}{\textbf{Personal Network}} \\
Network size & 5.61 (1.73) \\
Density & 0.67 (0.29) \\
Prop. of male alters & 0.54 (0.23) \\
Prop. of white alters & 0.71 (0.35) \\
Mean alter age & 45.66 (11.98) \\
Mean alter education & 4.07 (1.03) \\
Mean alter political leaning & 3.58 (1.28) \\
\midrule
\multicolumn{2}{l}{\textbf{Misinformation Susceptibility}} \\
Belief & 2.14 (0.85) \\
\hspace{.5cm}1. Ivermectin cures covid & 2.55 (1.80) \\
\hspace{.5cm}2. Vaccinated emit signals & 1.33 (1.08) \\
\hspace{.5cm}3. Sunscreen and cancer & 3.75 (1.56) \\
\hspace{.5cm}4. Bracelet cures diabetes & 1.49 (1.17) \\
\hspace{.5cm}5. Smoking prevents covid & 1.59 (1.33) \\
Sharing Intention & 2.18 (1.25) \\
\hspace{.5cm}1. Ivermectin cures covid & 2.44 (2.06) \\
\hspace{.5cm}2. Vaccinated emit signals & 1.57 (1.51) \\
\hspace{.5cm}3. Sunscreen and cancer & 3.58 (2.25) \\
\hspace{.5cm}4. Bracelet cures diabetes & 1.68 (1.60) \\
\hspace{.5cm}5. Smoking prevents covid & 1.63 (1.52) \\
\bottomrule
\end{tabular}
}
\caption{Descriptive statistics of public health survey.}
\label{tab:descriptive_health}
\end{table}
\rule{0pt}{17\baselineskip}

\pagebreak

\begin{table}[h]
\centering
\resizebox{0.4\textwidth}{!}{
\begin{tabular}{ll}
\toprule
\textbf{Variables} & \textbf{$M$ (SD) or \%} \\ 
\midrule
\multicolumn{2}{l}{\textbf{Demographics}} \\
Female & 59.95\% \\
White & 77.72\% \\
Age & 45.88 (13.52) \\
Education & 4.22 (1.51) \\
Income & 3.48 (1.59) \\
\midrule
\multicolumn{2}{l}{\textbf{Attitudinal/Behavioral}} \\
Political identification & 3.49 (1.27) \\
Systemic Processing & 3.94 (0.47) \\
Primary source: legacy media & 2.79 (1.74) \\
Primary source: social media & 4.39 (0.68) \\
No. of social media & 5.56 (2.42) \\
\midrule
\multicolumn{2}{l}{\textbf{Personal Network}} \\
Climate network size & 5.80 (3.46) \\
Climate alter prop. & 0.78 (0.23) \\
Prop of kin alters & 0.49 (0.28) \\
Mean tie strength & 4.03 (0.51) \\
Mutual awareness & 2.90 (0.84) \\
\midrule
\multicolumn{2}{l}{\textbf{Misinformation Susceptibility}} \\
Belief & 0.50 (0.36) \\
\hspace{.5cm}1. Stop climate change hype & 40.32\% \\
\hspace{.5cm}2. Paris Climate Treaty & 58.62\% \\
\hspace{.5cm}3. Mount Ena CO2 & 50.40\% \\
Sharing Intention & 0.23 (0.35) \\
\hspace{.5cm}1. Stop climate change hype & 23.61\% \\
\hspace{.5cm}2. Paris Climate Treaty & 24.14\% \\
\hspace{.5cm}3. Mount Ena CO2 & 20.42\% \\
\bottomrule
\end{tabular}
}
\caption{Descriptive statistics of climate change survey.}
\label{tab:descriptive_climate}
\end{table}

\rule{0pt}{17\baselineskip}

\pagebreak

\begin{table}[h]
\centering
\resizebox{0.4\textwidth}{!}{
\begin{tabular}{ll}
\toprule
\textbf{Variables} & \textbf{$M$ (SD) or \%} \\ 
\midrule
\multicolumn{2}{l}{\textbf{Demographics}} \\
Female & 46.75\% \\
Seoul metropolitan area & 83.42\% \\
Age & 47.21 (13.03) \\
Education & 5.60 (1.02) \\
Income & 4.97 (1.98) \\
\midrule
\multicolumn{2}{l}{\textbf{Attitudinal/Behavioral}} \\
Political identification & 4.80 (1.90) \\
Trust in science & 6.32 (1.57) \\
Social media use & 4.11 (1.16) \\
Health literacy & 3.26 (0.66) \\
Health media exposure & 2.81 (0.50) \\
\midrule
\multicolumn{2}{l}{\textbf{Personal Network}} \\
Network size & 4.86 (1.49) \\
Density & 0.72 (0.29) \\
Prop. of male alters & 0.24 (0.21) \\
Prop. of info alters & 0.50 (0.37) \\
Mean alter age & 46.32 (10.65) \\
Mean tie duration & 19.92 (11.92) \\
\midrule
\multicolumn{2}{l}{\textbf{Misinformation Susceptibility}} \\
Belief & 1.97 (0.80) \\
\hspace{.5cm}1. Gov't minimizes cases & 1.80 (0.87) \\
\hspace{.5cm}2. Gov't controls clinic info & 1.81 (0.83) \\
\hspace{.5cm}3. Masks and election fraud & 1.71 (0.88) \\
\hspace{.5cm}4. Mask shortage due to China & 2.29 (1.06) \\
\hspace{.5cm}5. Give masks away to China & 2.15 (1.00) \\
Sharing Intention & 1.27 (0.44) \\
\hspace{.5cm}1. Gov't minimizes cases & 1.19 (0.47) \\
\hspace{.5cm}2. Gov't controls clinic info & 1.30 (0.56) \\
\hspace{.5cm}3. Masks and election fraud & 1.43 (0.65) \\
\hspace{.5cm}4. Mask shortage due to China & 1.21 (0.49) \\
\hspace{.5cm}5. Give masks away to China & 1.21 (0.49) \\
\bottomrule
\end{tabular}
}
\caption{Descriptive statistics of pandemic politics survey.}
\label{tab:descriptive_politics}
\end{table}

\clearpage

\onecolumn
\appendix\section{Appendix B. Divergence and Correlation between Human and Simulated Susceptibility}

\begin{table*}[h]
\centering
\resizebox{.9\textwidth}{!}{
\begin{tabular}{lcccccc cccccc cccccc}
\toprule
 & \multicolumn{6}{c}{\textbf{Climate Change (US, 2025)}} & \multicolumn{6}{c}{\textbf{Public Health (US, 2023)}} & \multicolumn{6}{c}{\textbf{Pandemic Politics (KR, 2020)}} \\
\cmidrule(lr){2-7} \cmidrule(lr){8-13} \cmidrule(lr){14-19}
 & \multicolumn{3}{c}{\textbf{Misinfo Belief}} & \multicolumn{3}{c}{\textbf{Misinfo Sharing}} & \multicolumn{3}{c}{\textbf{Misinfo Belief}} & \multicolumn{3}{c}{\textbf{Misinfo Sharing}} & \multicolumn{3}{c}{\textbf{Misinfo Belief}} & \multicolumn{3}{c}{\textbf{Misinfo Sharing}} \\
\cmidrule(lr){2-4} \cmidrule(lr){5-7}
\cmidrule(lr){8-10} \cmidrule(lr){11-13}
\cmidrule(lr){14-16} \cmidrule(lr){17-19}
 & \textbf{JS $\downarrow$} & \textbf{EMD $\downarrow$} & \textbf{$\rho$ $\uparrow$} & \textbf{JS $\downarrow$} & \textbf{EMD $\downarrow$} & \textbf{$\rho$ $\uparrow$} & \textbf{JS $\downarrow$} & \textbf{EMD $\downarrow$} & \textbf{$\rho$ $\uparrow$} & \textbf{JS $\downarrow$} & \textbf{EMD $\downarrow$} & \textbf{$\rho$ $\uparrow$} & \textbf{JS $\downarrow$} & \textbf{EMD $\downarrow$} & \textbf{$\rho$ $\uparrow$} & \textbf{JS $\downarrow$} & \textbf{EMD $\downarrow$} & \textbf{$\rho$ $\uparrow$} \\
\midrule
\midrule
\textbf{Original Format} \\
\midrule
Reasoning Models\\
\addlinespace
 GPT-5.1 & .034 & .122 & .260 & .011 & \textbf{.039} & \underline{.170} & .171 & .119 & .189 & .023 & .060 & .249 & .027 & .093 & \textbf{.434} & \underline{.021} & .041 & \textbf{.102} \\
 GPT-5-mini & .074 & .208 & .240 & \textbf{.006} & \underline{.051} & .136 & .194 & .103 & .185 & .035 & .051 & .247 & .089 & .181 & .325 & .102 & .107 & -.021 \\
 Deepseek-V3 & \underline{.023} & \textbf{.056} & .283 & .035 & .169 & .135 & \underline{.047} & \underline{.053} & .172 & .051 & .063 & .216 & .133 & .120 & .186 & .195 & .112 & .001 \\
 Grok-4.1-fast & .054 & .087 & \textbf{.301} & .064 & .222 & .102 & \textbf{.041} & \textbf{.048} & .239 & \textbf{.010} & \textbf{.016} & \underline{.282} & \textbf{.021} & \textbf{.085} & \underline{.359} & .026 & .052 & \underline{.083} \\
 Olmo-3.1-32b-think & .042 & .159 & .279 & \underline{.009} & .068 & .145 & .277 & .148 & .045 & .040 & .067 & .171 & .241 & .198 & -.019 & .219 & .139 & -.042 \\
\midrule
Chat Models (CoT)\\
\addlinespace
 GPT-4.1-mini & .122 & .249 & .219 & .034 & .083 & .162 & .206 & .126 & .183 & .128 & .103 & .248 & .150 & .151 & .134 & .281 & .148 & -.072 \\
 Deepseek-V3 (nr) & \textbf{.021} & .104 & .289 & .034 & .123 & .138 & .056 & .055 & .230 & \underline{.016} & \underline{.023} & .219 & .095 & .114 & .242 & .032 & \underline{.037} & .002 \\
 Grok-4.1-fast (nr) & .027 & \textbf{.056} & .282 & .054 & .182 & \textbf{.171} & .087 & .104 & \underline{.243} & .017 & .051 & .225 & \underline{.026} & \underline{.087} & .235 & .045 & .070 & .002 \\
 Olmo-3.1-32b-inst & .125 & .256 & .158 & .053 & .133 & .136 & .220 & .136 & .187 & .080 & .072 & .254 & .195 & .157 & .114 & .272 & .154 & -.024 \\
\midrule
Chat Models (ZS)\\
\addlinespace
 GPT-4.1-mini & .076 & .176 & .261 & .049 & .104 & .169 & .421 & .160 & .084 & .203 & .150 & .178 & .239 & .210 & .101 & .143 & .069 & -.050 \\
 Deepseek-V3 (nr) & .031 & .134 & .286 & .034 & .083 & .131 & .088 & .086 & .176 & .031 & .038 & .215 & .137 & .128 & .286 & .025 & .054 & .008 \\
 Grok-4.1-fast (nr) & .034 & \underline{.060} & \underline{.296} & .058 & .203 & .143 & .240 & .132 & .195 & .060 & .075 & \textbf{.283} & .077 & .151 & .277 & \textbf{.020} & \textbf{.021} & .016 \\
 Olmo-3.1-32b-inst & .437 & .435 & .115 & .130 & .193 & .119 & .513 & .182 & .082 & .198 & .137 & .180 & .331 & .232 & .028 & .380 & .166 & .072 \\
\midrule
All Outputs Averaged & .037 & .114 & .293 & .032 & .085 & .154 & .113 & .100 & \textbf{.244} & .048 & .056 & .279 & .146 & .153 & .316 & .194 & .079 & .020 \\
\midrule
\midrule
\textbf{Alternative Ordering} \\
\midrule
Reasoning Models\\
\addlinespace
 GPT-5.1 & .032 & .124 & .256 & \underline{.009} & \textbf{.031} & .167 & .179 & .121 & .199 & .038 & .077 & .228 & \underline{.032} & \underline{.097} & \textbf{.441} & \underline{.014} & \underline{.033} & \textbf{.109} \\
 GPT-5-mini & .084 & .223 & .244 & \textbf{.005} & \underline{.041} & .181 & .197 & .103 & .210 & .034 & .052 & .203 & .092 & .187 & .293 & .088 & .101 & .011 \\
 Deepseek-V3 & .021 & .067 & .280 & .034 & .163 & .158 & \underline{.051} & \underline{.059} & .238 & .048 & .065 & .233 & .125 & .114 & .193 & .169 & .104 & -.041 \\
 Grok-4.1-fast & .031 & .065 & \underline{.300} & .051 & .202 & .147 & \textbf{.043} & \textbf{.048} & \underline{.242} & \textbf{.011} & \textbf{.017} & \underline{.256} & \textbf{.024} & \textbf{.096} & \underline{.360} & .021 & .046 & \underline{.070} \\
 Olmo-3.1-32b-think & .031 & .125 & .283 & .014 & .090 & .157 & .309 & .152 & .128 & .034 & .059 & .150 & .234 & .192 & -.023 & .204 & .130 & -.020 \\
\midrule
Chat Models (CoT)\\
\addlinespace
 GPT-4.1-mini & .129 & .259 & .227 & .024 & .069 & .160 & .177 & .121 & .194 & .090 & .083 & .219 & .141 & .161 & .120 & .242 & .127 & -.059 \\
 Deepseek-V3 (nr) & .023 & .107 & .249 & .021 & .124 & .129 & .074 & .062 & .215 & \underline{.019} & \underline{.023} & .226 & .097 & .111 & .238 & .050 & .037 & -.006 \\
 Grok-4.1-fast (nr) & \underline{.017} & \textbf{.040} & .262 & .047 & .179 & .146 & .082 & .097 & .222 & .019 & .045 & .236 & .055 & .140 & .246 & .046 & .053 & .022 \\
 Olmo-3.1-32b-inst & .130 & .266 & .213 & .063 & .144 & .165 & .206 & .132 & .139 & .031 & .045 & .255 & .189 & .167 & .084 & .221 & .135 & .061 \\
\midrule
Chat Models (ZS)\\
\addlinespace
 GPT-4.1-mini & .095 & .209 & .231 & .040 & .093 & \textbf{.224} & .392 & .156 & .041 & .147 & .116 & .245 & .290 & .226 & .084 & .072 & .052 & .008 \\
 Deepseek-V3 (nr) & .024 & .122 & .295 & .028 & .100 & .149 & .064 & .074 & .218 & .026 & .043 & .228 & .121 & .111 & .306 & .027 & .058 & .040 \\
 Grok-4.1-fast (nr) & \textbf{.012} & \underline{.065} & \textbf{.302} & .027 & .147 & .162 & .334 & .149 & .161 & .049 & .069 & .242 & .113 & .186 & .256 & \textbf{.012} & \textbf{.018} & -.003 \\
 Olmo-3.1-32b-inst & .475 & .444 & -.023 & .154 & .214 & \underline{.186} & .601 & .192 & .086 & .310 & .154 & .100 & .370 & .271 & .002 & .522 & .170 & -.005 \\
\midrule
All Outputs Averaged & .037 & .127 & .293 & .027 & .083 & .168 &
.117 & .101 & \textbf{.245} & .036 & .052 & \textbf{.260} &
.158 & .163 & .314 & .174 & .071 & .029 \\
\midrule
\midrule
\textbf{Composite Score Format} \\
\midrule
Reasoning Models\\
\addlinespace
 GPT-5.1 & .054 & .135 & .244 & .011 & \textbf{.039} & .161 & .334 & .152 & .176 & .107 & .131 & .225 & \underline{.033} & .104 & \textbf{.394} & \underline{.012} & \textbf{.028} & \textbf{.083} \\
 GPT-5-mini & .039 & .139 & .225 & .012 & .088 & .089 & .303 & .141 & .047 & .105 & .115 & .224 & .064 & .162 & .176 & .053 & .089 & .008 \\
 Deepseek-V3 & .020 & .077 & .276 & .025 & .138 & .124 & .101 & .071 & .203 & .028 & .032 & .238 & .091 & \textbf{.090} & .196 & .084 & .065 & .022 \\
 Grok-4.1-fast & .064 & .165 & .231 & \textbf{.008} & .055 & .126 & .122 & .094 & \underline{.222} & .045 & .086 & .269 & .040 & .128 & \underline{.297} & \textbf{.011} & \underline{.032} & \underline{.050} \\
 Olmo-3.1-32b-think & \underline{.017} & .088 & .289 & .018 & .100 & .115 & .326 & .156 & .140 & \textbf{.017} & \underline{.025} & .221 & .170 & .169 & -.029 & .194 & .155 & -.015 \\
\midrule
Chat Models (CoT)\\
\addlinespace
 GPT-4.1-mini & .060 & .156 & .261 & .020 & .053 & .154 & .242 & .133 & .182 & .127 & .086 & .286 & .109 & .138 & -.037 & .253 & .157 & -.011 \\
 Deepseek-V3 (nr) & .019 & .094 & .264 & .042 & .164 & .106 & .109 & .077 & .192 & \underline{.019} & \textbf{.020} & .235 & .103 & .118 & .094 & .045 & .050 & -.055 \\
 Grok-4.1-fast (nr) & .024 & .083 & .281 & .013 & .099 & .122 & \underline{.085} & \textbf{.067} & .169 & .023 & .053 & .281 & \textbf{.028} & \underline{.103} & .182 & .055 & .078 & -.005 \\
 Olmo-3.1-32b-inst & .039 & .144 & \textbf{.307} & \underline{.009} & \underline{.048} & \underline{.169} & .452 & .173 & .057 & .030 & .052 & .221 & .177 & .164 & -.004 & .211 & .132 & -.046 \\
\midrule
Chat Models (ZS)\\
\addlinespace
 GPT-4.1-mini & .053 & .127 & .293 & .041 & .070 & .134 & .328 & .141 & .116 & .144 & .088 & .288 & .254 & .206 & .029 & .188 & .112 & -.026 \\
 Deepseek-V3 (nr) & \textbf{.009} & \underline{.057} & .256 & .075 & .181 & \textbf{.669} & \textbf{.064} & \underline{.067} & .167 & .019 & .026 & .216 & .155 & .130 & .170 & .055 & .066 & -.009 \\
 Grok-4.1-fast (nr) & .025 & \textbf{.055} & \underline{.299} & .060 & .210 & .120 & .416 & .157 & .134 & .054 & .072 & \textbf{.323} & .101 & .187 & .170 & .052 & .038 & -.009 \\
 Olmo-3.1-32b-inst & .297 & .373 & .149 & .144 & .208 & .148 & .612 & .193 & .064 & .292 & .172 & .114 & .323 & .268 & -.020 & .420 & .135 & -.038 \\
\midrule
All Outputs Averaged & .030 & .109 & .290 & .020 & .086 & .115 & .210 & .130 & \textbf{.228} & .059 & .077 & \underline{.289} & .143 & .161 & .200 & .165 & .076 & .010 \\
\bottomrule
\end{tabular}
}
\caption{Distributional divergence and rank-order agreement between ground-truth and LLM-simulated misinformation susceptibility across survey domains, models, and prompting formats. Misinformation measures are min–max normalized.}
\label{tab:divergence}
\end{table*}

\clearpage

\twocolumn

\appendix\section{Appendix C. Simulation Prompt Format}

\appendix\subsection{C.1. Climate Change Simulation Prompt Format}

\begin{figure}[H]
    \centering
    \begin{tikzpicture}[font=\small]
        \node[
            draw,
            fill=gray!10,
            rounded corners,
            drop shadow={fill=black!30, shadow xshift=3pt, shadow yshift=-3pt},
            inner sep=10pt
        ] {
            \begin{minipage}{0.4\textwidth}
                \raggedright
                \begin{tabular}{l p{0.7\textwidth}}
                    \texttt{SYSTEM} &
                    This survey was conducted in the United States in Feb-Apr 2025.\newline
                    Suppose you are a survey participant described below, adopting their demographics, attitudes, behaviors, and personal network.\newline
                    Based ONLY on the profile, answer the given question.\newline
                    If the profile lacks information, make your best estimate from available signals; never reply with “unknown” or similar.\newline
                    Output format (JSON):\newline
                    \{``response'': ``integer''\}\newline
                    \\\
                    \texttt{INPUT} &
                    Participant profile:\newline
                    \{\textit{personal network information}\}\newline
                    \{\textit{demographic information}\}\newline
                    \{\textit{attitudes and behaviors}\}\newline
                    \newline
                    Consider the following claim:\newline
                    ``A single eruption of Mount Etna releases more carbon dioxide than all human activity combined.''\newline
                    \newline
                    Question: \newline
                    Please choose the option below that best describes your agreement with the above message. \newline 1. I agree with the information above \newline 2. I disagree with the information above. \newline
                    \textit{OR} \newline
                    Question:\newline 
                    Please choose the option below that best describes your intention to share the above message with other people.\newline
                    1. I am likely to share it with others. \newline
                    2. I am unlikely to share it with others.
                \end{tabular}
            \end{minipage}
        };
    \end{tikzpicture}
    \caption{Example prompt format used to simulate climate change survey participants’ misinformation belief or sharing intention. Belief and sharing questions were queried separately using identical system prompts, participant profiles, and claims.}
    \label{fig:prompt_survey_sim_climate}
\end{figure}

\rule{0pt}{10\baselineskip}

\pagebreak
\appendix\subsection{C.2. Pandemic Politics Simulation Prompt Format}

\begin{figure}[H]
    \centering
    \begin{tikzpicture}[font=\small]
        \node[
            draw,
            fill=gray!10,
            rounded corners,
            drop shadow={fill=black!30, shadow xshift=3pt, shadow yshift=-3pt},
            inner sep=10pt
        ] {
            \begin{minipage}{0.4\textwidth}
                \raggedright
                \begin{tabular}{l p{0.7\textwidth}}
                    \texttt{SYSTEM} &
                    This survey was conducted in South Korea in May 2020.\newline
                    Suppose you are a survey participant described below, adopting their demographics, attitudes, behaviors, and personal network.\newline
                    Based ONLY on the profile, answer the given question.\newline
                    If the profile lacks information, make your best estimate from available signals; never reply with “unknown” or similar.\newline
                    Output format (JSON):\newline
                    \{``response'': ``integer''\}\newline
                    \\\
                    \texttt{INPUT} &
                    Participant profile:\newline
                    \{\textit{personal network information}\}\newline
                    \{\textit{demographic information}\}\newline
                    \{\textit{attitudes and behaviors}\}\newline
                    \newline
                    Consider the following claim:\newline
                    ``Purchasing public masks at a pharmacy will lead to leakage of personal information which will be used in election fraud.''\newline
                    \newline
                    Question:\newline
                    Above is information related to the coronavirus and coronavirus disease (COVID-19). To the best of your knowledge, how accurate do you think the following statements are? (1=Not at all accurate; 4=Very accurate) \newline
                    \textit{OR} \newline
                    Question:\newline
                    Would you be likely to share the following information related to the coronavirus and coronavirus disease (COVID-19) with others via social media such as Facebook or Twitter, or via messaging apps such as KakaoTalk, LINE, or WhatsApp? Please indicate how likely you are to share it. (1=No, I would not share it; 2=I would probably share it; 3=Yes, I would share it)
                \end{tabular}
            \end{minipage}
        };
    \end{tikzpicture}
    \caption{Example prompt format used to simulate pandemic politics survey participants’ misinformation belief or sharing intention. Belief and sharing questions were queried separately using identical system prompts, participant profiles, and claims.}
    \label{fig:prompt_survey_politics}
\end{figure}

\rule{0pt}{6\baselineskip}

\appendix\section{Appendix D. Original Profile Format}
\appendix\subsection{D.1. Public Health Survey Participant Profile}

\begin{figure}[H]
    \centering
    \begin{tikzpicture}[font=\scriptsize]
        \node[draw, fill=mygray, rounded corners, drop shadow={fill=black!30, shadow xshift=1pt, shadow yshift=-3pt, opacity=0.5}, inner sep=10pt] {
            \begin{minipage}{0.45\textwidth}
                        \textbf{Demographics}\par
                        \begin{itemize}
                            \item Gender: Female
                            \item Age: 42
                            \item Race/Ethnicity: White
                            \item Education: Bachelor's degree
                            \item Household income: \$25,000 to \$49,999
                        \end{itemize}
                                        \textbf{Attitudes and Behaviors}\par
                \textit{Political Leaning}
                \begin{itemize} 
                    \item Here is a 7-point scale on which the political views that people might hold are arranged from extremely liberal (1) to extremely conservative (7). Where would you place yourself on this scale?: 5
                \end{itemize}
                \textit{Trust in Science}\par
                Please state how much you trust:\par
                \begin{itemize}
                    \item Scientists in the US: A lot
                    \item Science: A lot
                    \item Scientists to find out accurate information about the world: A lot
                    \item Scientists working in colleges or universities benefiting the public: A lot
                    \item Scientists working in colleges or universities being honest about who is paying for their work: A lot
                    \item Scientists working for companies making medicines or agricultural products benefiting the public: A lot
                    \item Scientists working for companies making medicines or agricultural products being honest about who is paying for their work: A lot
                \end{itemize}
                \vspace{0.5em}
                \textit{Health Literacy}\par
                \begin{itemize}
                    \item How often do you have someone help you read hospital materials: Never
                    \item How confident are you filling out medical forms by yourself: Extremely
                    \item How often do you have problems learning about your medical condition because of difficulty understanding written information: Never
                    \item How often do you have a problem understanding what is told to you about your medical condition: Never
                \end{itemize}
                \vspace{0.5em}
                \textit{Social Media Use}\par
                \begin{itemize}
                    \item How often do you use social media platform (e.g., Facebook, Instagram, Twitter) to get health-related news and information?: Once a day
                \end{itemize}
                \vspace{0.5em}
                \textit{Health Information Use (past 30 days)}\par
                How often have you done each of the following in the past 30 days?\par
                \begin{itemize}
                    \item Read health information on the Internet: Once per week
                    \item Read about health issues in newspapers or general magazines: Once per week
                    \item Watched special health segments of television newscasts: Everyday
                    \item Watched television programs (other than news) which address health issues or focus on doctors or hospitals: Everyday
                \end{itemize}
            \end{minipage}
            };
    \end{tikzpicture}
    \caption{Illustrative example of demographic and attitudinal/behavioral feature blocks included in public health survey participant profiles. Responses are replaced with fictitious examples to maintain anonymity.}
    \label{fig:demo_prompt_health}
\end{figure}

\pagebreak

\begin{figure}[H]
    \centering
    \begin{tikzpicture}[font=\small]
        \node[draw, fill=mygray, rounded corners, drop shadow={fill=black!30, shadow xshift=1pt, shadow yshift=-3pt, opacity=0.5}, inner sep=10pt] {
            \begin{minipage}{0.4\textwidth}
                        \textbf{Personal Network}
                        \newline \newline
                        \textit{Contacts}\par
                        \begin{itemize}
                            \item You listed 3 regular contact(s): pp, cc, sh
                            \item You discuss health issues with: pp
                        \end{itemize}
                        \vspace{0.5em}
                        \textit{Mutual Awareness}\par
                        \begin{itemize}
                            \item pp knows cc
                            \item cc knows pp, sh
                            \item sh knows cc
                        \end{itemize}
                        \vspace{0.5em}
                        \textit{Contact Profiles}\par
                        \vspace{0.5em}
                        pp
                        \begin{itemize}
                            \item Gender: Other
                            \item Age: 26
                            \item Race: Asian
                            \item Education: Bachelor's Degree
                            \item Political leaning: Liberal
                        \end{itemize}
                        cc
                        \begin{itemize}
                            \item Gender: Male
                            \item Age: 30
                            \item Race: Asian
                            \item Education: Master's Degree
                            \item Political leaning: Slightly liberal
                        \end{itemize}
                        sh
                        \begin{itemize}
                            \item Gender: Female
                            \item Age: 15
                            \item Race: Asian
                            \item Education: Less than high school
                            \item Political leaning: Moderate
                        \end{itemize}                        
                        \par
                        \vspace{0.5em}  
            \end{minipage}
            };
    \end{tikzpicture}
    \caption{Illustrative example of a network feature block included in public health survey participant profiles. Responses are replaced with fictitious examples to maintain anonymity.}
    \label{fig:net_prompt_health}
\end{figure}

\clearpage

\appendix\subsection{D.2. Climate Change Survey Participant Profile}

\begin{figure}[H]
    \centering
    \begin{tikzpicture}[font=\scriptsize]
        \node[
            draw,
            fill=mygray,
            rounded corners,
            drop shadow={fill=black!30, shadow xshift=1pt, shadow yshift=-3pt, opacity=0.5},
            inner sep=10pt
        ] {
            \begin{minipage}{0.45\textwidth}
            
            \textbf{Demographics}\par
            \begin{itemize}
                \item Gender: Female
                \item Age: 42
                \item Race: White
                \item Education: High school diploma or GED
                \item Household income: Less than \$30{,}000
            \end{itemize}
            
            \textbf{Attitudes and Behaviors}\par
            \vspace{0.5em}
            
            \textit{Political Leaning}\par
            \begin{itemize}
                \item In general, you consider yourself to be: Conservative
            \end{itemize}
            
            \vspace{0.5em}
            \textit{Systematic Processing}\par
            \begin{itemize}
                \item After I encounter information about complex societal issues, I am likely to stop and think about it.: Agree
                \item If I need to act on complex societal issues, the more viewpoints I get the better.: Strongly agree
                \item It is important for me to interpret information about complex societal issues in a way that applies directly to my life.: Neither agree nor disagree
                \item After thinking about complex societal issues, I have a broader understanding.: Agree
                \item When I encounter information about complex societal issues, I read or listen to most of it, even though I may not agree with its perspective.: Agree
            \end{itemize}
            
            \vspace{0.5em}
            \textit{Primary Information Source}\par
            \begin{itemize}
                \item How do you primarily access information about climate change?: Social media platforms
            \end{itemize}
            
            \vspace{0.5em}
            \textit{Social Media Use}\par
            \begin{itemize}
                \item Which social media platforms do you have accounts on and log on at least past month? Please select all that apply.: Facebook, Instagram, Pinterest, Reddit, X/Twitter, YouTube
            \end{itemize}
            
            \end{minipage}
        };
    \end{tikzpicture}
    \caption{Illustrative example of demographic and attitudinal/behavioral feature blocks included in climate change survey participant profiles. Responses are replaced with fictitious examples to maintain anonymity.}
    \label{fig:demo_prompt_climate}
\end{figure}

\rule{0pt}{17\baselineskip}

\pagebreak
\begin{figure}[H]
    \centering
    \begin{tikzpicture}[font=\small]
        \node[
            draw,
            fill=mygray,
            rounded corners,
            drop shadow={fill=black!30, shadow xshift=1pt, shadow yshift=-3pt, opacity=0.5},
            inner sep=10pt
        ] {
            \begin{minipage}{0.4\textwidth}
            
            \textbf{Personal Network}
            \newline \newline
            
            \textit{Contacts}\par
            \begin{itemize}
                \item You listed 3 contact(s): pp, cc, sh
                \item You discuss climate change with 2 contact(s): pp, sh
            \end{itemize}
            
            \vspace{0.5em}
            \textit{Perceived Mutual Awareness}\par
            \begin{itemize}
                \item To the best of your knowledge, please indicate the degree to which you think they know one another: Most of them know each other
            \end{itemize}
            
            \vspace{0.5em}
            \textit{Climate Discussant Profiles}\par
            \vspace{0.5em}
            
            pp
            \begin{itemize}
                \item Relationship: Acquaintance
                \item Closeness (1--5 scale): 4
            \end{itemize}
            
            sh
            \begin{itemize}
                \item Relationship: Family
                \item Closeness (1--5 scale): 5
            \end{itemize}
            
            \vspace{0.5em}
            \end{minipage}
        };
    \end{tikzpicture}
    \caption{Illustrative example of a network feature block included in climate change survey participant profiles. Responses are replaced with fictitious examples to maintain anonymity.}
    \label{fig:net_prompt_climate}
\end{figure}

\clearpage

\appendix\subsection{D.3. Pandemic Politics Survey Participant Profile}

\begin{figure}[H]
    \centering
    \begin{tikzpicture}[font=\scriptsize]
        \node[
            draw,
            fill=mygray,
            rounded corners,
            drop shadow={fill=black!30, shadow xshift=1pt, shadow yshift=-3pt, opacity=0.5},
            inner sep=10pt
        ] {
            \begin{minipage}{0.45\textwidth}
            
            \textbf{Demographics}\par
            \begin{itemize}
                \item Gender: Male
                \item Age: 35
                \item Region: Capital Region
                \item Education: Associate degree
                \item Monthly household income: 1--2 million KRW
            \end{itemize}
            
            \textbf{Attitudes and Behaviors}\par
            \vspace{0.5em}
            
            \textit{Political Leaning}\par
            \begin{itemize}
                \item On an 11-point scale of political (ideological) orientation, where would you place yourself? (0 = Extremely liberal, 5 = Moderate, 10 = Extremely conservative): 5
            \end{itemize}
            
            \vspace{0.5em}
            \textit{Trust in Science}\par
            \begin{itemize}
                \item When seeking information about controversial science-related issues, how much do you trust scientists? (0 = Not at all, 10 = Completely): 7
            \end{itemize}
            
            \vspace{0.5em}
            \textit{Health Literacy}\par
            \begin{itemize}
                \item How often do you have someone help you read hospital materials?: Sometimes
                \item How often do you have problems learning about your medical condition because of difficulty understanding written information?: Occasionally
                \item How often do you have a problem understanding what is told to you about your medical condition?: Sometimes
                \item How confident are you filling out medical forms by yourself?: Quite a bit
            \end{itemize}
            
            \vspace{0.5em}
            \textit{Social Media Use}\par
            \begin{itemize}
                \item How often do you usually use social networking services such as Facebook, KakaoStory, Instagram, Naver Band, or Twitter (X)?: Several times a day
                \item How often do you usually use instant mobile messaging services such as KakaoTalk, LINE, or Telegram?: Once a day
            \end{itemize}
            
            \vspace{0.5em}
            \textit{Health Information Use}\par
            In the past month, how often have you encountered news reports or information about health and medical issues through the following sources?\par
            \begin{itemize}
                \item Daily newspapers or magazines: Not at all
                \item Television: Rarely
                \item Online news: Frequently
                \item Health- or medical-specialty websites: Rarely
            \end{itemize}

            \end{minipage}
        };
    \end{tikzpicture}
    \caption{Illustrative example of demographic and attitudinal/behavioral feature blocks included in political survey participant profiles. Responses are replaced with fictitious examples to maintain anonymity.}
    \label{fig:demo_prompt_politics}
\end{figure}

\rule{0pt}{9\baselineskip}

\begin{figure}[H]
    \centering
    \begin{tikzpicture}[font=\small]
        \node[
            draw,
            fill=mygray,
            rounded corners,
            drop shadow={fill=black!30, shadow xshift=1pt, shadow yshift=-3pt, opacity=0.5},
            inner sep=10pt
        ] {
            \begin{minipage}{0.4\textwidth}
            
            \textbf{Personal Network}\par
            \vspace{0.5em}
            
            \textit{Contacts}\par
            \begin{itemize}
                \item You listed 3 contact(s): ㅍㅍ, ㅊㅊ, ㅅㅎ
            \end{itemize}
            
            \vspace{0.5em}
            \textit{Mutual Awareness}\par
            \begin{itemize}
                \item ㅍㅍ knows ㅊㅊ
                \item ㅊㅊ knows ㅍㅍ, ㅅㅎ
                \item ㅅㅎ knows ㅊㅊ
            \end{itemize}
            
            \vspace{0.5em}
            \textit{Contact Profiles}\par
            \vspace{0.5em}
            
            ㅍㅍ
            \begin{itemize}
                \item Gender: Female
                \item Age: 26
                \item Relationship duration (years): 2.1
                \item Mainly provides information support: Yes
            \end{itemize}
            
            ㅊㅊ
            \begin{itemize}
                \item Gender: Male
                \item Age: 30
                \item Relationship duration (years): 3.1
                \item Mainly provides information support: Yes
            \end{itemize}
            
            ㅅㅎ
            \begin{itemize}
                \item Gender: Female
                \item Age: 15
                \item Relationship duration (years): 15
                \item Mainly provides information support: No
            \end{itemize}
            
            \vspace{0.5em}
            \end{minipage}
        };
    \end{tikzpicture}
    \caption{Illustrative example of a network feature block included in political survey participant profiles. Responses are replaced with fictitious examples to maintain anonymity.}
    \label{fig:net_prompt_politics}
\end{figure}

\clearpage
\appendix\section{Appendix E. Composite Scores Profile Format}

\appendix\subsection{E.1. Public Health Survey Participant Profile}

\begin{figure}[h]
    \centering
    \begin{tikzpicture}[font=\scriptsize]
        \node[
            draw,
            fill=mygray,
            rounded corners,
            drop shadow={fill=black!30, shadow xshift=1pt, shadow yshift=-3pt, opacity=0.5},
            inner sep=10pt
        ] {
            \begin{minipage}{0.45\textwidth}
            
            \textbf{Personal Network}\par
            \begin{itemize}
                \item Total number of alters: 7
                \item Network density: 0.48
                \item Alters by gender: Female 43\%, Male 57\%
                \item Alters by race/ethnicity: White 43\%, Black 29\%, Latino 14\%, Native 14\%
                \item Average age of alters: 29
                \item Average education of alters: Associate degree
                \item Average political leaning of alters: Liberal
            \end{itemize}
            
            \textbf{Demographics}\par
            \begin{itemize}
                \item Gender: Female
                \item Age: 27
                \item Race: White
                \item Education: Bachelor's degree
                \item Yearly household income: \$25{,}000 to \$49{,}999
            \end{itemize}
            
            \textbf{Attitudes and Behaviors}\par
            \begin{itemize}
                \item Political Leaning (1--7 scale): 1
                \item Trust in Science (1--4 scale): 3.0
                \item Health Literacy (1--5 scale): 5.0
                \item Social Media Use (1--6 scale): 4.0
                \item Health Information Use (1--4 scale): 3.2
            \end{itemize}
            
            \end{minipage}
        };
    \end{tikzpicture}
    \caption{Illustrative example of demographic, network, and attitudinal features included in public health survey participant profiles. Responses are replaced with fictitious examples to maintain anonymity.}
    \label{fig:demo_prompt_comp_health}
\end{figure}

\rule{0pt}{22\baselineskip}

\pagebreak

\appendix\subsection{E.2. Climate Change Survey Participant Profile}

\begin{figure}[h]
    \centering
    \begin{tikzpicture}[font=\scriptsize]
        \node[
            draw,
            fill=mygray,
            rounded corners,
            drop shadow={fill=black!30, shadow xshift=1pt, shadow yshift=-3pt, opacity=0.5},
            inner sep=10pt
        ] {
            \begin{minipage}{0.4\textwidth}
            
            \textbf{Personal Network}\par
            \begin{itemize}
                \item Number of climate alters: 3
                \item Share of all named alters: 100\%
                \item Kinship share of climate alters: 33\%
                \item Mean tie strength with climate alters (1--5 scale): 3.8
                \item Mutual awareness among all named alters (1--4 scale): 3
            \end{itemize}
            
            \textbf{Demographics}\par
            \begin{itemize}
                \item Gender: Male
                \item Age: 54
                \item Race: White
                \item Education: Bachelor's Degree
                \item Household income: Less than \$30{,}000
            \end{itemize}
            
            \textbf{Attitudes and Behaviors}\par
            \begin{itemize}
                \item Political Leaning: Somewhat conservative
                \item Systematic processing (1--5 scale): 3.8
            \end{itemize}
            
            \vspace{0.5em}
            \textit{Primary Climate Change Information Source}\par
            \begin{itemize}
                \item Legacy media: No
                \item Online media: Yes
            \end{itemize}
            
            \vspace{0.5em}
            \textit{Social Media Use}\par
            \begin{itemize}
                \item Number of social media platforms used (min 0; max 11): 2
            \end{itemize}
            
            \end{minipage}
        };
    \end{tikzpicture}
    \caption{Illustrative example of demographic, network, and attitudinal features included in climate change survey participant profiles. Responses are replaced with fictitious examples to maintain anonymity.}
    \label{fig:demo_prompt_comp_climate}
\end{figure}

\rule{0pt}{22\baselineskip}

\pagebreak

\appendix\subsection{E.3. Pandemic Politics Survey Participant Profile}

\begin{figure}[h]
    \centering
    \begin{tikzpicture}[font=\scriptsize]
        \node[
            draw,
            fill=mygray,
            rounded corners,
            drop shadow={fill=black!30, shadow xshift=1pt, shadow yshift=-3pt, opacity=0.5},
            inner sep=10pt
        ] {
            \begin{minipage}{0.45\textwidth}
            
            \textbf{Personal Network (the people you discuss important matters)}\par
            \begin{itemize}
                \item Total number of alters: 2
                \item Network density: 0.0
                \item Portion of female alters: 100.0\%
                \item Portion of male alters: 0.0\%
                \item Portion of alters providing information support: 100.0\%
                \item Average age of alters: 39.0
                \item Average duration of relationship with alters (years): 15.0
            \end{itemize}
            
            \textbf{Demographics}\par
            \begin{itemize}
                \item Gender: Male
                \item Age: 43
                \item Region: Non-capital Region
                \item Education: Bachelor's degree
                \item Monthly household income: 5--6 million KRW
            \end{itemize}
            
            \textbf{Attitudes and Behaviors}\par
            \begin{itemize}
                \item Political Leaning (0--10 scale): 5.0
                \item Trust in Science (0--10 scale): 6.0
                \item Health Literacy (1--5 scale): 4.0
                \item Social Media Use (1--6 scale): 3.0
                \item Health Information Use (1--4 scale): 3.1
            \end{itemize}
            
            \end{minipage}
        };
    \end{tikzpicture}
    \caption{Illustrative example of demographic, network, and attitudinal features included in political survey participant profiles. Responses are replaced with fictitious examples to maintain anonymity.}
    \label{fig:demo_prompt_comp_politics}
\end{figure}

\rule{0pt}{22\baselineskip}

\clearpage
\onecolumn

\appendix\section{Appendix F. Results from Alternative Prompts}\label{sec:survey}
\appendix\subsection{F.1. Block Removal Analysis}

\begin{figure*}[h]
    \centering
    \includegraphics[width=.9\textwidth]{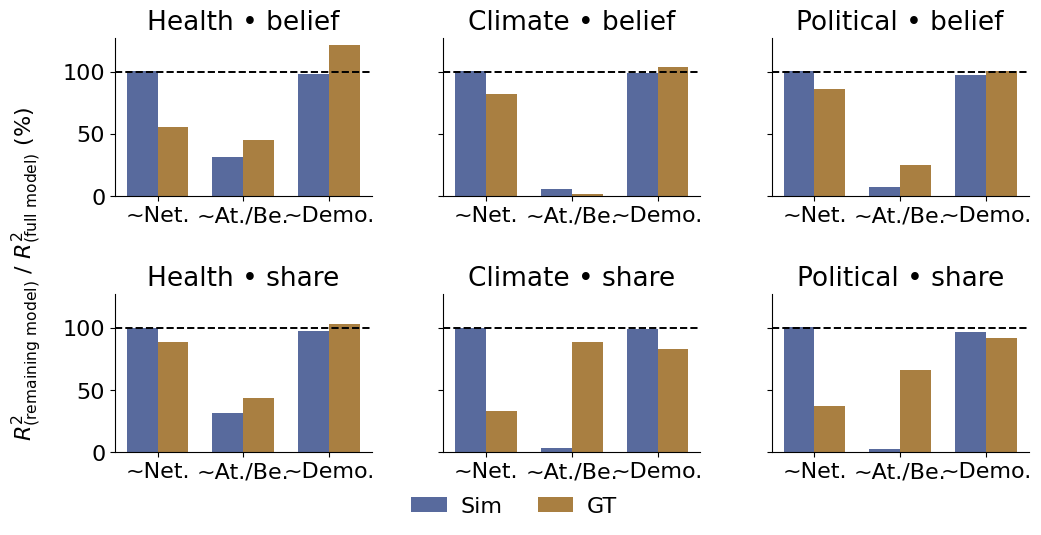}
    \caption{Relative explained variance after removing each feature block. This serves as an ablation-style analysis of feature contributions. Bars show the proportion of predictive power retained after block removal, $R^2_{\mathrm{block\ removed}}/R^2_{\mathrm{full\ model}}$ (in \%), for simulated (\textcolor{myblue}{Blue}) and ground-truth (\textcolor{myyellow}{Orange}) outcomes. All $R^2$ values are 5-fold cross-validated. $\sim$Net.: network block removed; $\sim$At./Be.: attitudinal/behavioral block removed; $\sim$Demo.: demographics block removed.}
    \label{fig:block_removal_original}
\end{figure*}

\clearpage

\begin{figure}[h]
    \centering
    \includegraphics[width=.8\textwidth]{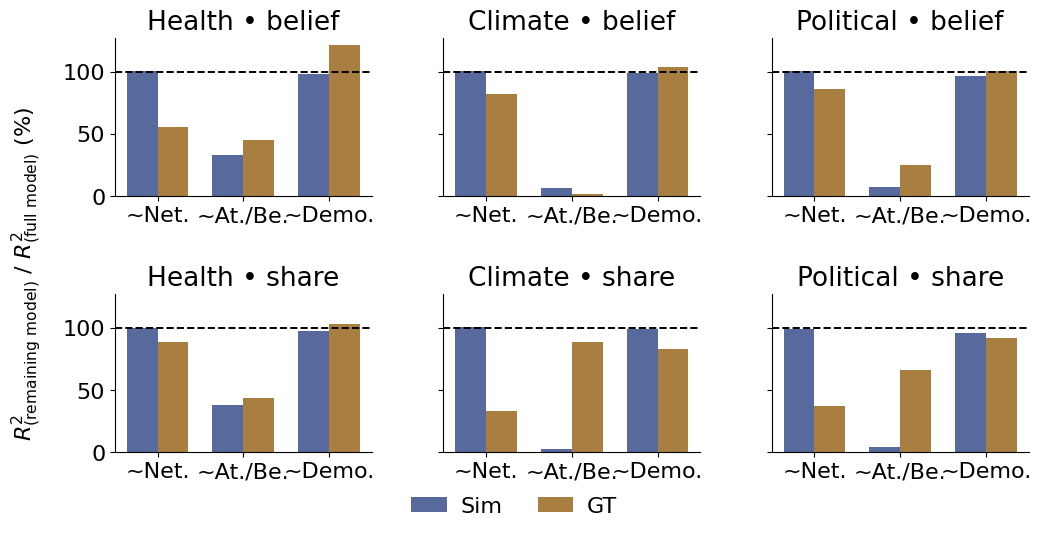}
    \caption{Relative explained variance after removing each feature block. Alternative profile ordering.}
    \label{fig:block_removal_altseq}
\end{figure}

\begin{figure}[h]
    \centering
    \includegraphics[width=.8\textwidth]{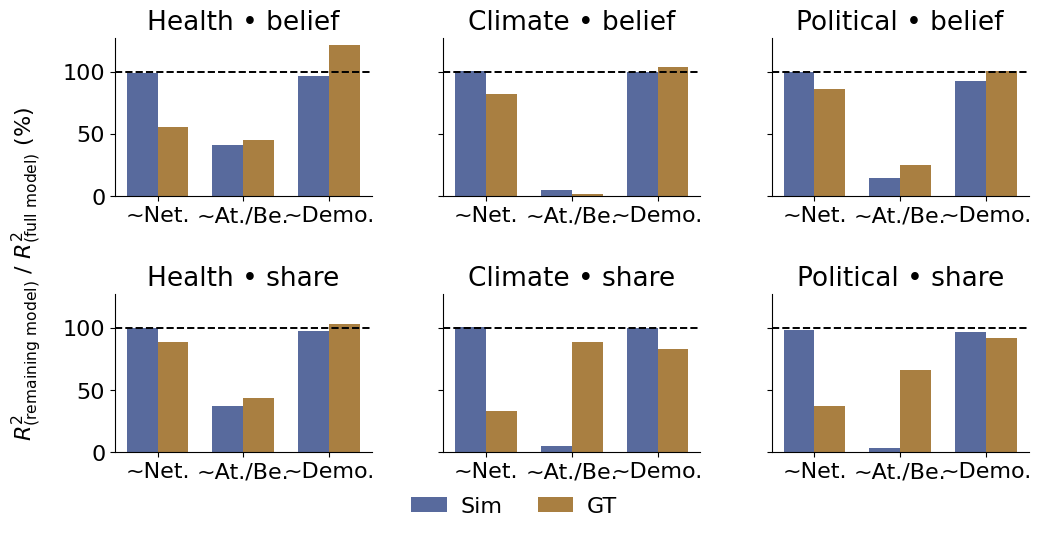}
    \caption{Relative explained variance after removing each feature block. Composite score profile format.}
    \label{fig:block_removal_composite}
\end{figure}

\rule{0pt}{10\baselineskip}

\clearpage
\appendix\subsection{F.2. Interaction Analysis}

\begin{figure*}[h]
    \centering
    \includegraphics[width=.7\textwidth]{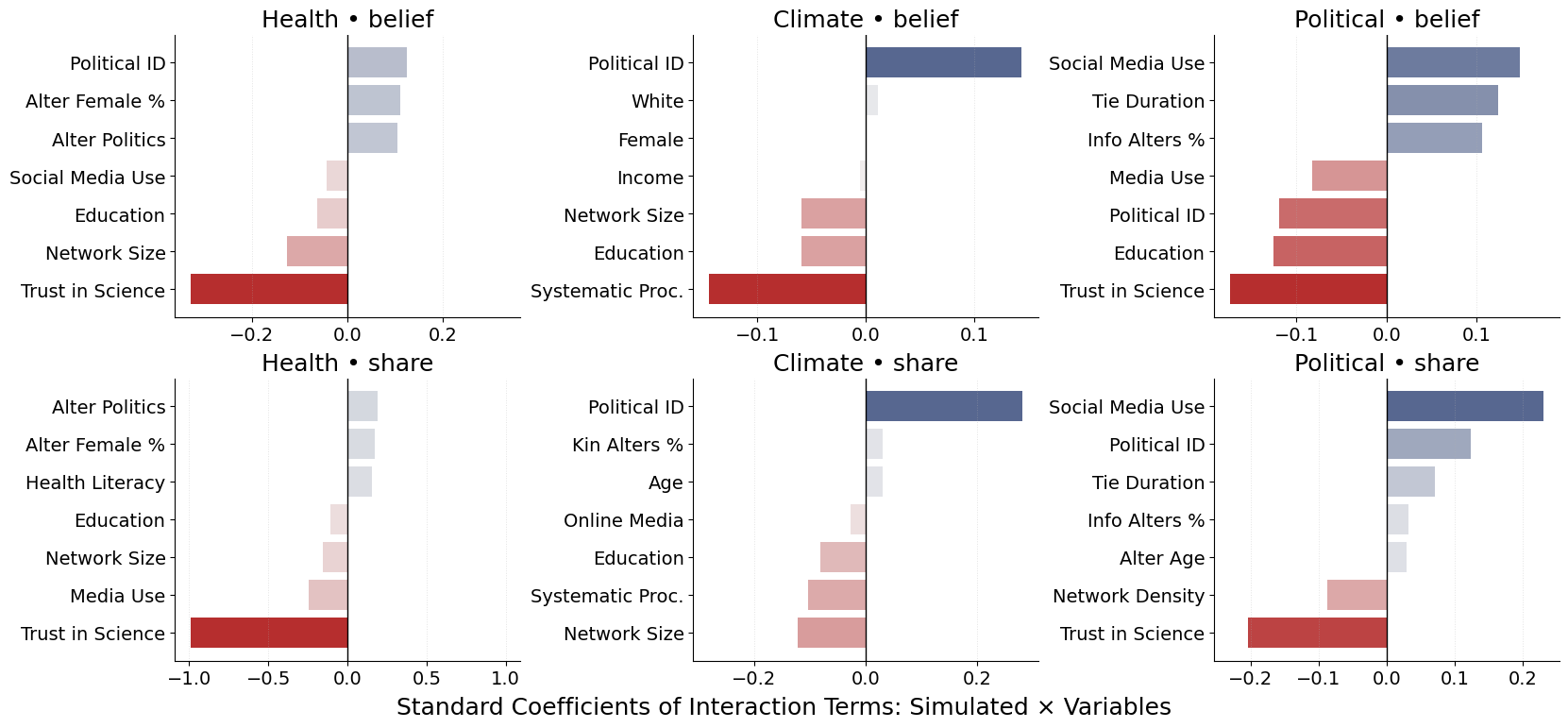}
    \caption{Standardized coefficients for simulation-by-feature interaction terms (Simulated $\times$ Variables) across domains and outcomes. Alternative profile ordering.}
    \label{fig:sim_interactions_altseq}
\end{figure*}

\begin{figure*}[h]
    \centering
    \includegraphics[width=.7\textwidth]{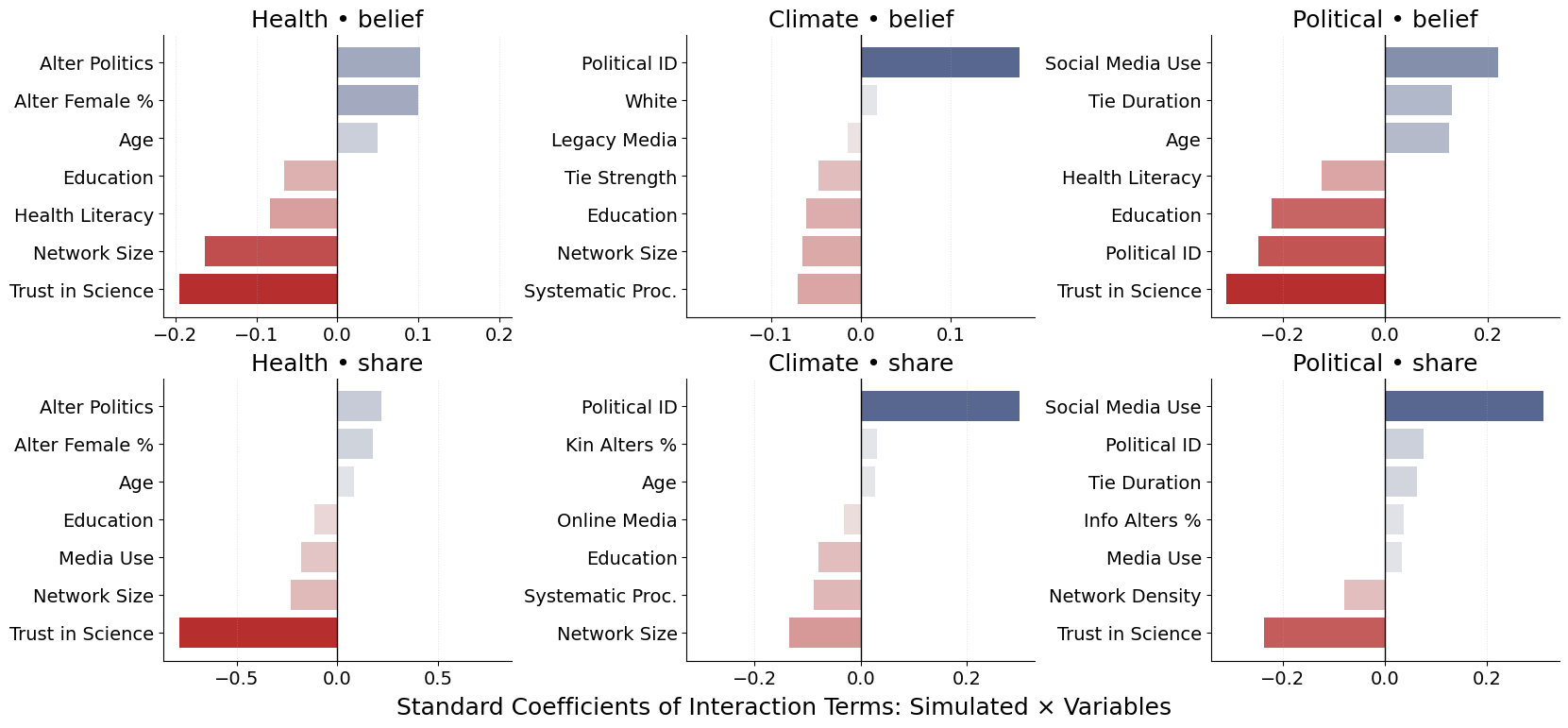}
    \caption{Standardized coefficients for simulation-by-feature interaction terms (Simulated $\times$ Variables) across domains and outcomes. Composite score profile format.}
    \label{fig:sim_interactions_composite}
\end{figure*}

\clearpage
\twocolumn
\appendix\section{Appendix G. Procedure of Model Reasoning and Training Corpus Analysis}\label{sec:alt}

We examine two plausible, non-exclusive mechanisms that may help contextualize the distorted simulation results: (i) differential salience of predictors in model-generated reasoning traces and (ii) uneven representation of predictor–misinformation associations in training data, as visualized in Figure~\ref{fig:hypothesized_mechanism}.

We first examine model reasoning processes. Our dataset includes reasoning chains produced by open-weight reasoning models (DeepSeek-V3 and OLMo-3.1-32B-Think) and chat models prompted with explicit chain-of-thought (CoT) instructions (GPT-4.1-mini, DeepSeek-V3, Grok-4.1-fast in non-reasoning mode, and OLMo-3.1-32B-Instruct). Across belief/sharing intention items and survey domains, this yields 42,584 unique reasoning chains. 

To identify which profile features were invoked during model reasoning, we conducted a post-hoc extraction analysis using three independent LLMs (GPT-4.1-mini, DeepSeek-V3, and Grok-4.1-fast). Each model was prompted to identify variables that appeared most salient for supporting the reasoning, restricted to features explicitly mentioned or directly implied in the reasoning chain, and to return the result as a Python list (prompt shown in Figure~\ref{fig:important_variables_prompt}). Agreement across the three model-based extractions returned an average pairwise Jaccard similarity of .750.

To reduce the influence of ambiguous or weakly implied variables, we adopted a conservative inclusion rule: a variable was retained as important for a given reasoning chain only when all three extraction models independently selected it. Under this unanimity setting, each reasoning chain yielded an average of 4.07 retained variables (\textit{SD} = 1.90).

As a robustness check, two independent human raters, one of whom was an author, manually annotated 100 randomly sampled reasoning chains using the same decision rules as the prompt (Figure~\ref{fig:important_variables_prompt}). Agreement between model-retained and human annotations was moderate (average pairwise Jaccard similarity = .628), indicating moderate alignment. Accordingly, we interpret the extracted variables as coarse indicators of reasoning salience rather than precise representations of underlying cognitive processes.

Next, we examine whether similar patterns emerge from LLM training data. While the training corpora of major proprietary models are not publicly accessible, we analyze open-source training data made available by the Allen Institute for AI (AI2) using OLMoTrace, a tool that retrieves up to 10 training-data spans most relevant to a given query. For each profile feature, we construct four query templates combining the variable name with misinformation-related concepts: (i) \{\textit{variable}\} and misinformation, (ii) misinformation and \{\textit{variable}\}, (iii) \{\textit{variable}\} and disinformation, and (iv) disinformation and \{\textit{variable}\}. To improve coverage, each variable is queried using three paraphrased name variants; the complete list of variable names and aliases is provided in Table~\ref{tab:var_alias}. Retrieved spans are deduplicated and post-processed by retaining only those in which all query terms explicitly appear in the text. This procedure yields 188 unique spans. 

To characterize the direction of association implied by each retrieved span, we further processed the text using three independent LLMs (GPT-4.1-mini, DeepSeek-V3, and Grok-4.1-fast). Each model was prompted to classify whether the association between the queried variable and misinformation susceptibility reflected in the span was positive, negative, or neutral (prompt shown in Figure~\ref{fig:corpus_prompt}). For example, given a span mentioning trust in science, models assessed whether the text implied a higher, lower level of susceptibility, or no clear directional association.

Agreement across the three model-based annotations was moderate (Fleiss’ $\kappa = .580$). To minimize the influence of ambiguous or weakly specified associations, we adopted a strict coding rule: a span was labeled as positive or negative only when all three models reached unanimous agreement; spans lacking unanimity were treated as neutral and excluded from directional interpretation.

As an additional robustness check, two independent human raters, one of whom was an author, manually annotated the retrieved training corpus spans using the same labeling scheme. Agreement between model-consensus labels and human annotations was high (Fleiss’ $\kappa = .634$), suggesting a fair amount of overlap in inferred associations. We treat these annotations as indicative rather than definitive and use them to support comparisons of associative patterns rather than causal claims.

\begin{figure*}[h]
    \centering
    \begin{tikzpicture}[font=\small]
        \node[
            draw,
            fill=gray!10,
            rounded corners,
            drop shadow={fill=black!30, shadow xshift=3pt, shadow yshift=-3pt},
            inner sep=10pt
        ] {
            \begin{minipage}{\textwidth}
                \raggedright

                You are given a model’s reasoning and a list of candidate variables that may explain the conclusion of the reasoning. \newline

                \textbf{Model reasoning:} \newline
                \{model reasoning\}
                
                \vspace{0.75em}
                \textbf{Candidate variables:}\par
                
                \textit{(1) Demographics}\par
                gender, age, race, education, income, region
                
                \vspace{0.5em}
                \textit{(2) Attitudinal}\par
                political leaning, trust in science, health literacy
                
                \vspace{0.5em}
                \textit{(3) Behavioral}\par
                social media use, legacy media use, online media use
                
                \vspace{0.5em}
                \textit{(4) Network Characteristics}\par
                network size, relationship with alters, issue-specific discussants, tie strength, tie duration, education level of alters, political leaning of alters, age of alters, gender of alters, race of alters, information support from alters, network density, mutual awareness
                
                \vspace{0.75em}
                \textbf{Task:} \newline 
                Select the set of variables that DIRECTLY SUPPORTS the conclusion according to the reasoning above.

                \vspace{0.75em}
                \textbf{Rules:} \newline 
                Only include variables that are explicitly implied or directly used in the reasoning. \\
                Do not include redundant or weakly related variables.
                
                \vspace{0.5em}
                \textbf{Output:} \newline 
                Return ONLY valid JSON with exactly these keys: \newline
                \{
                ``reasoning'': ``Brief evidence-based rationale (1–3 sentences).'',
                 ``label'': [List of selected variable names from the candidate variables]
                \}
                
            \end{minipage}
        };
    \end{tikzpicture}
    \caption{Prompt used to identify key explanatory variables from model-generated reasoning.}
    \label{fig:important_variables_prompt}
\end{figure*}

\begin{figure*}[h]
    \centering
    \begin{tikzpicture}[font=\small]
        \node[
            draw,
            fill=gray!10,
            rounded corners,
            drop shadow={fill=black!30, shadow xshift=3pt, shadow yshift=-3pt},
            inner sep=10pt
        ] {
            \begin{minipage}{\textwidth}
                \raggedright
                
                You are an expert researcher analyzing text data.\par
                
                \vspace{0.75em}
                \textbf{Text to Analyze:} \newline ``\{text\}''\par
                
                \vspace{0.75em}
                \textbf{Task:} \newline Based ONLY on the text above, determine the DIRECTION OF ASSOCIATION between the variable ``\{variable\_name\}'' and the concept/outcome ``\{target\}'' as described in the text, in a statistical, correlational, or causal sense.\par
                
                \vspace{0.75em}
                \textbf{Interpretation:} \newline ``Positive'' means the text implies that higher/more of ``\{variable\_name\}'' is associated with higher/more of ``\{target\}'' in belief or sharing. ``Negative'' means the text implies that higher/more of ``\{variable\_name\}'' is associated with lower/less of ``\{target\}'' in belief or sharing. ``Neutral'' means the text does not clearly specify a direction (only mentions both, is descriptive without linking them, is ambiguous, mixed/conditional with no net direction, or no association is stated).\par
                
                \vspace{0.5em}
                For the following variables, always interpret the association relative to the specified reference category: \\Gender: Interpret the variable as a binary indicator of being female (female = 1, not female = 0). \\Race: Interpret the variable as a binary indicator of being White (White = 1, non-White = 0). \\Region: Interpret the variable as a binary indicator of living in a Metropolitan Area (metro = 1, non-metro = 0). \\Political leaning: Interpret the variable as a binary indicator of being conservative (conservative = 1, not conservative = 0).\par
                
                \vspace{0.75em}
                \textbf{Rules:} \newline Only consider belief in ``\{target\}'' or sharing tendency. If the text only states that they are related/associated without direction, return ``Neutral''. If the text states both directions in different contexts without a clear overall direction, return ``Neutral''. Do NOT use sentiment or moral approval/disapproval to decide the label.\par
                
                \vspace{0.75em}
                \textbf{Output:} \newline
                Return ONLY valid JSON with exactly these keys:
                \{
                ``reasoning'': ``Brief evidence-based rationale (1--3 sentences).'',
                 ``label'': ``Positive'' \textbar\ ``Neutral'' \textbar\ ``Negative''
                \}
                
            \end{minipage}
        };
    \end{tikzpicture}
    \caption{Prompt used for direction-of-association annotation.}
    \label{fig:corpus_prompt}
\end{figure*}

\begin{table*}[b]
\centering
\begin{tabularx}{\textwidth}{p{0.15\textwidth} X p{0.20\textwidth} X}
\toprule
\textbf{Variable} & \textbf{Alias} & \textbf{Variable} & \textbf{Alias} \\
\midrule

\multirow{3}{*}{gender} & gender & \multirow{3}{*}{network size} & network size \\
 & sex &  & group size \\
 & gender identity &  & social reach \\
\addlinespace

\multirow{3}{*}{age} & age & \multirow{3}{*}{relationship with alters} & relationship with alters \\
 & years &  & ties with contacts \\
 & respondent age &  & connections with others \\
\addlinespace

\multirow{3}{*}{race} & race & \multirow{3}{*}{climate discussant} & climate discussant \\
 & ethnicity &  & climate interlocutor \\
 & ancestry &  & climate facilitator \\
\addlinespace

\multirow{3}{*}{education} & education & \multirow{3}{*}{tie strength} & tie strength \\
 & education level &  & relationship closeness \\
 & educational attainment &  & connection intensity \\
\addlinespace

\multirow{3}{*}{income} & income & \multirow{3}{*}{tie duration} & tie duration \\
 & earnings &  & connection length \\
 & household income &  & bond longevity \\
\addlinespace

\multirow{3}{*}{region} & region & \multirow{3}{*}{education level of alters} & education level of alters \\
 & area &  & alters education \\
 & locale &  & alters educational attainment \\
\addlinespace

\multirow{3}{*}{political leaning} & political leaning & \multirow{3}{*}{political leaning of alters} & political leaning of alters \\
 & political orientation &  & alters political orientation \\
 & political stance &  & alters political ideology \\
\addlinespace

\multirow{3}{*}{trust in science} & trust in science & \multirow{3}{*}{age of alters} & age of alters \\
 & faith in science &  & alters' ages \\
 & confidence in science &  & network member age \\
\addlinespace

\multirow{3}{*}{health literacy} & health literacy & \multirow{3}{*}{gender of alters} & gender of alters \\
 & health knowledge &  & alter gender \\
 & health understanding &  & alters' sex \\
\addlinespace

\multirow{3}{*}{social media use} & social media use & \multirow{3}{*}{race of alters} & race of alters \\
 & social media engagement &  & alter race \\
 & online platform usage &  & alters race \\
\addlinespace

\multirow{3}{*}{legacy media use} & legacy media use & \multirow{3}{*}{info. support from alters} & information support from alters \\
 & traditional media consumption &  & information support from network members \\
 & mainstream media exposure &  & informational assistance from close contacts \\
\addlinespace

\multirow{3}{*}{online media use} & online media use & \multirow{3}{*}{network density} & network density \\
 & digital media consumption &  & connection density \\
 & internet media engagement &  & tie density \\
\addlinespace

 &  & \multirow{3}{*}{mutual awareness} & mutual awareness \\
 &  &  & shared awareness \\
 &  &  & reciprocal awareness \\

\bottomrule
\end{tabularx}
\caption{Variable names and aliases used to query OLMoTrace.}
\label{tab:var_alias}
\end{table*}

\end{document}